\documentclass[11pt, a4paper]{article}
\pdfoutput=1
\usepackage{jcappub}

\usepackage{booktabs}
\usepackage{multirow}
\usepackage{amssymb}
\usepackage[export]{adjustbox}
\usepackage{floatrow}
\newfloatcommand{capbtabbox}{table}[][3.5in]
\newfloatcommand{capbfigbox}{figure}[][2.3in]
\usepackage{blindtext}
\usepackage{amsmath}
\usepackage{booktabs}
\usepackage{aas_macros}

\newcommand{\ie}{i.e.~}

\def\lsim{\mathrel{\raise.3ex\hbox{$<$\kern-.75em\lower1ex\hbox{$\sim$}}}}
\def\gsim{\mathrel{\raise.3ex\hbox{$>$\kern-.75em\lower1ex\hbox{$\sim$}}}}

\begin{document}

\hspace*{110mm}{\large \tt FERMILAB-PUB-17-203-A}
\vskip 0.2in

\title{High-Energy Gamma Rays and Neutrinos from Nearby Radio Galaxies}

\author{Carlos Blanco$^{a,b}$}\note{ORCID: http://orcid.org/0000-0001-8971-834X}
\emailAdd{carlosblanco2718@uchicago.edu}
\author{and Dan Hooper$^{b,c,d}$}\note{ORCID: http://orcid.org/0000-0001-8837-4127}
\emailAdd{dhooper@fnal.gov}

\affiliation[a]{University of Chicago, Department of Physics, Chicago, IL 60637}
\affiliation[b]{University of Chicago, Kavli Institute for Cosmological Physics, Chicago, IL 60637}
\affiliation[c]{Fermi National Accelerator Laboratory, Center for Particle Astrophysics, Batavia, IL 60510}
\affiliation[d]{University of Chicago, Department of Astronomy and Astrophysics, Chicago, IL 60637}

\abstract{Multi-messenger data suggest that radio galaxies (\ie non-blazar active galaxies) are a well-motivated class of sources for the diffuse flux of high-energy neutrinos reported by the IceCube Collaboration. In this study, we consider the gamma-ray spectrum observed from four nearby radio galaxies (Centaurus A, PKS 0625-35, NGC 1275 and IC 310) and constrain the intensity and spectral shape of the emission injected from these sources, accounting for the effects of attenuation and contributions from electromagnetic cascades (initiated both within the radio galaxy itself and during extragalactic propagation). Assuming that this gamma-ray emission is generated primarily through the interactions of cosmic-ray protons with gas, we calculate the neutrino flux predicted from each of these sources. Although this scenario is consistent with the constraints published by the IceCube and ANTARES Collaborations, the predicted fluxes consistently fall within an order of magnitude of the current point source sensitivity. The prospects appear very encouraging for the future detection of neutrino emission from the nearest radio galaxies.}

\maketitle

\section{Introduction}

Since the IceCube Collaboration announced their discovery of a diffuse flux of high-energy astrophysical neutrinos~\cite{Aartsen:2013bka}, a great deal of effort has been directed toward identifying the sources of these intriguing particles. Among other possibilities, it has long been speculated that observable fluxes of high-energy neutrinos could potentially be generated by gamma-ray bursts~\cite{Waxman:1997ti,Rachen:1998ir,Guetta:2003wi}, active galaxies~\cite{Stecker:1991vm,Halzen:1997hw,Atoyan:2001ey,Murase:2015ndr}, and star-forming galaxies~\cite{Loeb:2006tw} (for reviews, see Refs.~\cite{Becker:2007sv,Halzen:2002pg}). When scrutinized in the light of multi-messenger data, however, many of these proposed neutrino sources do not fare well. In particular, IceCube has not detected any statistically significant correlations in time or direction between their events and observed gamma-ray bursts, all but ruling out the possibility that such sources are responsible for the measured neutrino flux~\cite{Abbasi:2012zw,Aartsen:2016qcr} (very low-luminosity gamma-ray bursts may be able to evade this constraint, however~\cite{Murase:2013ffa,Tamborra:2015qza,Senno:2015tsn,Tamborra:2015fzv}). Furthermore, an analysis of IceCube data has not revealed any correlations with the members of the Fermi Gamma-Ray Space Telescope's blazar catalog, leading to the conclusion that no more than 20\% of IceCube's flux can originate from this class of active galaxies~\cite{Glusenkamp:2015jca} (see also Ref.~\cite{Ahlers:2014ioa}). 

Supernova and hypernova remnants in star-forming and starburst galaxies may also generate significant fluxes of high-energy neutrinos, with spectra that could potentially extend up to energies of $\sim$\,0.1--1~PeV. Interactions generating neutrinos in these sources would inevitably produce a comparable flux of gamma-rays, however, leading to the production of electromagnetic cascades and thus contributing to the GeV-TeV extragalactic gamma-ray background. In light of recent evidence that the extragalactic gamma-ray background is dominated by blazars at energies above 50 GeV~\cite{TheFermi-LAT:2015ykq}, it now appears that the sources of IceCube's neutrino flux must exhibit a significantly harder spectral index than that observed from star-forming galaxies. More specifically, the authors of Refs.~\cite{Bechtol:2015uqb,Murase:2015xka} argue that the 10-100 TeV neutrino flux observed by IceCube cannot be generated by this class of sources (although they may be able to explain the flux observed above $\sim$100 TeV~\cite{Bechtol:2015uqb,Chakraborty:2016mvc,Xiao:2016rvd}). Similar, but more general, arguments can be extended to show that any class of sources which produces IceCube's neutrino flux through proton-proton interactions must exhibit a hard spectral index, $<$\,2.1--2.2, and will inevitably contribute non-negligibly to the high-energy gamma-ray background measured by Fermi~\cite{Murase:2013rfa,Hooper:2016jls} (see also Refs.~\cite{Chang:2016ljk,Murase:2015xka,Kalashev:2014vra}).

As sources such as gamma-ray bursts and blazars have become disfavored, non-blazar active galaxies (those with jets not aligned in the direction of the observer) have become an increasingly attractive class of sources for IceCube's high-energy neutrinos~\cite{Hooper:2016jls,Murase:2015ndr,Giacinti:2015pya,Tjus:2014dna} (see also Refs.~\cite{Murase:2014foa,Kimura:2014jba,Murase:2016gly,Sahu:2012wv,Marinelli:2014mva,Fraija:2017jok,Fraija:2012na}).  In this paper, we will refer to this class of objects as radio galaxies, which includes both Fanaroff-Riley (FR) Type I and Type II galaxies, which are generally thought to be the misaligned counterparts of BL Lac objects and flat spectrum radio quasars, respectively. This conclusion is further strengthened by the observed characteristics of the isotropic gamma-ray background (IGRB), as measured by the Fermi Gamma-Ray Space Telescope~\cite{Ackermann:2014usa}. In particular, it has been shown in recent studies that radio galaxies~\cite{Hooper:2016gjy} and star-forming galaxies~\cite{Linden:2016fdd} (see also Refs.~\cite{Tamborra:2014xia,Ackermann:2012vca,Stecker:2010di,DiMauro:2013xta}) together provide the dominant contributions to the IGRB (smaller but potentially non-negligible contributions also come from blazars~\cite{Cuoco:2012yf,Harding:2012gk,Ajello:2011zi,Ajello:2013lka,Stecker:2010di}, as well as from galaxy clusters~\cite{Zandanel:2014pva}, propagating ultra-high energy cosmic rays~\cite{Taylor:2015rla,Ahlers:2011sd}, and perhaps even annihilating dark matter particles~\cite{Ackermann:2015tah,DiMauro:2015tfa,Ajello:2015mfa,Cholis:2013ena}). In quantitative terms, Ref.~\cite{Hooper:2016gjy} concludes that unresolved radio galaxies account for no less than 59\% of the IGRB photons above 1 GeV, while Ref.~\cite{Linden:2016fdd} finds that star-forming galaxies are responsible for at least 24\% of the IGRB intensity above 1 GeV (each at the 2$\sigma$ confidence level). Together, these analyses strongly support a scenario in which the IGRB is dominated by these two source classes, and largely by radio galaxies at energies above $\sim$10 GeV. This conclusion is also consistent with the findings of previous work~\cite{Fornasa:2015qua,DiMauro:2016cbj,DiMauro:2015tfa,Ajello:2015mfa,Cholis:2013ena,Cavadini:2011ig,Siegal-Gaskins:2013tga}, including studies based on cross-correlations of the IGRB with multi-wavelength data~\cite{Xia:2015wka,Cuoco:2015rfa,Shirasaki:2014noa,Shirasaki:2015nqp,Ando:2015bva}.

To unambiguously demonstrate that radio galaxies are responsible for IceCube's observed neutrino spectrum, it will be essential to identify high-energy neutrinos from individual radio galaxies. In this paper, we consider the gamma-ray observations of a number of nearby radio galaxies (focusing on Centaurus A, PKS 0625-35, NGC 1275 and IC 310), using data from both Fermi and ground-based gamma-ray telescopes. Assuming that the observed gamma-ray emission is generated through proton-proton interactions, we use the measured gamma-ray spectra of these sources to characterize and constrain the cosmic-rays that are generated by these sources, as well as the energy densities of radiation and magnetic fields present in these environments. We then calculate the neutrino spectra anticipated from these radio galaxies and assess the prospects for IceCube and future neutrino telescopes to detect neutrinos from these individual sources.

\section{Gamma-Ray Observations of Nearby Radio Galaxies}

\begin{table*}
\label{Table1}
\begin{tabular}{|c|c|c|c|c|c|}
\hline 
Galaxy  & Distance  & Radius  & Luminosity & GeV Variability & VHE Variability \tabularnewline
\hline 
\hline 
Milky Way  & --  & 17 kpc & $2\times10^{10} \, L_{\odot}$ & -- & -- \tabularnewline
\hline 
Centaurus A  & 4  Mpc & 15 kpc  & $2.5\times10^{10}\, L_{\odot}$ & No (59.3) & No~\cite{Aharonian:2009xn} \tabularnewline
\hline 
PKS 0625-35  & 228 Mpc  & 23 kpc  & $4.4\times10^{11}\, L_{\odot}$ & No (58.0) & No~\cite{Dyrda:2015hxa} \tabularnewline
\hline 
NGC 1275  & 75 Mpc  & 24 kpc & $1.5\times10^{11}\, L_{\odot}$ & Yes (622.2) & Yes~\cite{Ahnen:2016qkt,2017ATel.9931....1M,2016ATel.9690....1M} \tabularnewline
\hline 
IC 310  & 80 Mpc  & 27 kpc & $7.88\times10^{10}\, L_{\odot}$ & No (38.7) & Yes~\cite{Aleksic:2013bya}
\tabularnewline
\hline 
\end{tabular}
\caption{The approximate distance, disk radius, and luminosity of the radio galaxies considered in this study. Values for the Milky Way are also shown for comparison. In the last two columns, we list whether each source has been reported to be variable at GeV energies by Fermi, and at very-high energies by VERITAS, HESS or MAGIC. In the case of Fermi, we provide (in parentheses) the reported variability index; any value greater than 59.78 (72.44) is less than 10\% (1\%) likely to be a steady source~\cite{TheFermi-LAT:2015hja}.}
\end{table*}

The Fermi Collaboration has reported the detection of gamma-ray emission from a number of radio galaxies, including 14 which are listed in the Third Fermi Gamma-Ray Source Catalog (3FGL)~\cite{TheFermi-LAT:2015hja}. More recently, statistically significant emission has also been identified from the radio galaxies 3C 212, 3C 411 and B3 0309+411B, each utilizing publicly available Fermi data~\cite{Hooper:2016gjy}. In addition, ground-based telescopes have detected very high-energy gamma-ray emission from a number of radio galaxies, including Centaurus A (by HESS)~\cite{Aharonian:2009xn}, M~87 (by VERITAS, MAGIC and HESS)~\cite{Acciari:2008ah,Acciari:2009rs,Galante:2009ie,:2012uma}, PKS 0625-35 (by HESS)~\cite{Dyrda:2015hxa}, NGC 1275 (by VERITAS and MAGIC)~\cite{Galante:2009ie,Aleksic:2013kaa} and 3C 310 (by MAGIC)~\cite{Galante:2009ie}.

The mechanisms which generate the observed gamma-ray emission from these sources is not yet entirely clear. While some radio galaxies exhibit considerable variability, thus favoring leptonic (\ie inverse Compton) models, other radio galaxies are not observed to be variable, and may produce their gamma-ray emission through hadronic processes (\ie pion production). If hadronic processes are responsible for a significant fraction of the gamma-ray emission from these sources, radio galaxies would also be predicted to generate a flux of high-energy neutrinos similar to that observed by IceCube~\cite{Hooper:2016jls} (see also Refs.~\cite{Murase:2016gly,Sahu:2012wv,Marinelli:2014mva,Fraija:2017jok,Fraija:2012na}). Furthermore, if the accelerated cosmic rays have a spectrum that extends with an unbroken index of approximately $\sim$\,1.8, radio galaxies could accommodate the observed fluxes of gamma rays and neutrinos, while simultaneously explaining the spectrum of the ultra-high energy cosmic rays.

In this study, we focus on those radio galaxies that have been observed at TeV energies by ground-based gamma-ray telescopes (for a review of gamma-ray emission from non-blazar active galaxies, see Ref.~\cite{Rieger2017}). In particular, we will consider the following list of radio galaxies (see also Table~1):

\begin{itemize}

\item{{\bf Centaurus A} is the nearest radio galaxy in our sample (4 Mpc), and is comparable in
both size and luminosity to the Milky Way. This galaxy does not show significant
evidence of variability at energies above a couple of GeV~\cite{Aharonian2009,Rieger2017}.}

\item{{\bf PKS 0625-35} is the most distant radio galaxy in our sample (228 Mpc), but with a luminosity that is more than an order of magnitude larger than that of the Milky Way. PKS 0625-35 does not exhibit any significant evidence of variability~\cite{Dyrda:2015hxa,Rieger2017}.}

\item{{\bf NGC 1275} is a radio galaxy located near the center of the Perseus cluster. It presents a
peculiar morphology consisting of extended filament-like gas
structures which are likely caused by interactions with intracluster
gas. Furthermore, there is a foreground object in the line-of-sight to NGC 1275 which may be interacting with
the central galaxy. The so-called ``high-velocity system'' is thought to be a late-type galaxy whose morphology has been disturbed through interactions with NGC 1275. This high-velocity system is less than 57 kpc from the center
of NGC 1275 and spans $\sim25$ kpc~\cite{yu2015high,gillmon2004}. Fermi reports a high degree of variability in the emission from NGC 1275~\cite{brown2011high}. At very high-energies, however, MAGIC reports only limited evidence of such variability~\cite{Ahnen:2016qkt,Conselice2001,Rieger2017,aleksic2014contemporaneous}.}

\item{{\bf IC 310} is another radio galaxy located in the Perseus cluster. Although observations by Fermi have revealed no signs of variability, significant variability is evident in the TeV range, as observed by MAGIC~\cite{aleksic2014}. It has been suggested that the very high-energy emission from IC 310 can be separated into two distinct states; a highly variable ``high state'' and a steady baseline ``low state''~\cite{aleksic2014}.  While the overall intensity varies by a factor of $\sim$7 between these two states, no distinction in the spectral index is observed.}

\end{itemize}

Due to its strong and rapid variability in the very high-energy band~\cite{abramowski2012,albert2008,aliu2012}, we do not consider further the radio galalxy M~87 in this study.

\section{Electromagnetic Cascades in Radio Galaxies}

At very high-energies, gamma rays can be significantly attenuated through their scattering with the optical, infrared and microwave radiation fields present in radio galaxies (as well as during cosmological propagation). In this section we will describe the calcluation of the electromagnetic cascades that result from such interactions (for other recent work on this subject, see Refs.~\cite{Murase:2011yw,Murase:2012xs,Murase:2011cy,Murase:2012df,Berezinsky:2016feh}). The optical depth associated with interactions in an isotropic radiation field is given by:
\begin{eqnarray}
\tau(E_{\gamma})&=&  \int \int   \sigma_{\gamma\gamma}(E_{\gamma},\epsilon) \, \frac{dn}{d\epsilon}(\epsilon,r) \, d\epsilon \, dr \\
&\approx&  l \int  \sigma_{\gamma\gamma}(E_{\gamma},\epsilon) \, \frac{dn}{d\epsilon}(\epsilon) \, d\epsilon, \nonumber
\end{eqnarray}
where $E_{\gamma}$ is the energy of the gamma ray, $\epsilon$ is the energy of the target photon, $\sigma_{\gamma\gamma}$ is the total pair-production cross section, and $dn(\epsilon, r)/d\epsilon$ is the differential number
density of target photons at a location $r$. In the second line of this expression, we have integrated over the trajectory of the gamma ray, yielding a result in terms of the average number density of target photons, $dn(\epsilon)/d\epsilon$, and a characteristic size for the extent of the target radiation, $l$. The total cross section for pair-production is well approximated by the following expression~\cite{aharonian1983}:
\begin{eqnarray}
\sigma_{\gamma\gamma}(s)&=&\frac{3\sigma_{T}}{2s^{2}}\left(s-1+\frac{1}{2s}-\frac{\ln s}{2}+\ln2\right) \\
&\times& \bigg(\ln\left(\sqrt{s}+\sqrt{s-1}\right)+\frac{\left(\ln s\right)^{2}}{4}-\frac{\left(\ln\left(\sqrt{s}+\sqrt{s-1}\right)\right)^{2}}{2}+\frac{\ln2 \, \ln s}{2}- \sqrt{s^2-s} \bigg), \nonumber
\end{eqnarray}
where $s=E_{\gamma} \epsilon/m_{e}^{2}$ and $\sigma_T$ is the Thomson cross section.

The spectrum of electrons and positrons that are generated from these interactions is given by~\cite{aharonian1983}:
\begin{equation}
\frac{dN_e}{dE_{e}}(E_{e}) =l\iint\frac{dN_{\gamma}}{dE_{\gamma}}(E_{\gamma}) \,\frac{dn}{d\epsilon}(\epsilon)\,\frac{d\sigma_{\gamma\gamma}}{dE_{e}}(\epsilon,E_{\gamma},E_{e})    \,d\epsilon \, dE_{\gamma},
\end{equation}
where $dN_{\gamma}/dE_{\gamma}$ is the spectrum of gamma rays injected by the radio galaxy in question, and $d\sigma_{\gamma \gamma}/dE_e$ is the differential pair-production cross section, given by:
\begin{eqnarray}
\frac{d\sigma_{\gamma\gamma}}{dE_{e}}(\epsilon,E_{\gamma},E_{e})&=&\frac{3\sigma_{T}m_{e}^{4}}{32\epsilon^{2}E_{\gamma}^{3}} \, \bigg[ \frac{4E_{\gamma}^{2}}{(E_{\gamma}-E_{e}) E_{e}} \, 
\ln \bigg(\frac{4\epsilon E_{e}(E_{\gamma}-E_{e})}{m_{e}^{2}E_{\gamma}}\bigg) -
\frac{8\epsilon E_{\gamma}}{m_{e}^{2}} \nonumber \\
&+& \bigg(\frac{2E_{\gamma}^{2} (2\epsilon E_{\gamma}-m_{e}^{2})}{(E_{\gamma}-E_{e})E_{e}m_{e}^{2}}\bigg)
-\bigg(1-\frac{m_{e}^{2}}{\epsilon E_{\gamma}}\bigg) \, 
\frac{E_{\gamma}^{4}}{(E_{\gamma}-E_{e})^{2}E_{e}^{2}}\bigg) \bigg].
\end{eqnarray}

The energetic electron-positron pairs that are generated through these interaction will interact with the radiation and magnetic fields through inverse Compton scattering and synchrotron processes, respectively. Over the course of losing a quantity of energy, $\Delta E_e$, an electron or positron of energy $E_e$ will generate the following spectrum of inverse Compton emission:
\begin{eqnarray}
\frac{dN_{\gamma}}{dE_{\gamma}}(E_{\gamma}, E_e) &=& A(E_e,\Delta E_e) \, f_{\rm ICS}(E_e) \, l_{e} \int    \frac{dn}{d\epsilon}(\epsilon) \, \frac{d\sigma_{ICS}}{dE_{\gamma}}(\epsilon,E_{\gamma},E_{e}) \,   d\epsilon,
\end{eqnarray}
where $A$ is set by requirement $\Delta E_e = \int dE_{\gamma} \, E_{\gamma} \, dN_{\gamma}/dE_{\gamma}$, $f_{\rm ICS}(E_e)$ is the fraction of the electron or positron's energy losses that go to inverse Compton scattering (as opposed to synchrotron), and the differential cross section for inverse Compton scattering is given by~\cite{aharonian1981}:
\begin{eqnarray}
\frac{d\sigma_{ICS}}{dE_{\gamma}}(\epsilon,E_{\gamma},E_{e})&=&\frac{3\sigma_{T}m_{e}^{2}}{4\epsilon E_{e}^{2}}\,
\bigg[ 1+\bigg( \frac{z^{2}}{2(1-z)}\bigg)+\bigg(\frac{z}{\beta (1-z)}\bigg)-\bigg(\frac{2z^{2}}{\beta^{2}(1-z)}\bigg) \nonumber \\
&-&\bigg(\frac{z^{3}}{2\beta (1-z)^{2}}\bigg)-\bigg(\frac{2z}{\beta (1-z)}\bigg)\ln\bigg(\frac{\beta (1-z)}{z}\bigg) \bigg],
\end{eqnarray}
where $z \equiv E_{\gamma}/E_{e}$ and $\beta \equiv 4\epsilon E_{e}/m_{e}^{2}$.

High-energy electrons also lose energy through their interactions with magnetic fields, via synchrotron emission. For a randomly oriented
magnetic field of uniform stength, the synchrotron energy loss rate is given by:
\begin{equation}
\frac{dE_e}{dt} = \frac{4\sigma_{T}}{3m_e^2 c^3}U_{B} E_e^2, 
\end{equation}
where $U_B = B^2/8 \pi$ is the energy density of the magnetic field. In the Thomson limit, the corresponding energy loss rate resulting from inverse Compton scattering is given by:
\begin{equation}
\frac{dE_e}{dt} = \frac{4\sigma_{T}}{3m_e^2 c^3}U_{\rm rad} E_e^2.
\end{equation}
For the problem at hand, however, we will often find ourselves well outside of the Thomson regime, and Klein-Nishina suppression will play an important role. For a radiation spectrum that consists of a combination of blackbodies, the energy loss rate from inverse Compton scattering can be written as follows~\cite{schlickeiser2010}: 
\begin{equation}
\frac{dE_e}{dt} = \frac{4\sigma_{T}}{3m_e^2 c^3} \sum_i U_{i, \rm rad} E_e^2 \, \bigg(\frac{\gamma_{i, k}^2}{\gamma_{i, k}^2+\gamma^2}\bigg),
\end{equation}
where $\gamma=E_e/m_e$, $U_{i, \rm rad}$ is the energy density in the $i$th component of radiation, and
\begin{equation}
\gamma_{i, k}=\frac{3\sqrt{5}m_{e}c^{2}}{8\pi k_{b}T_i},
\end{equation}
where $T_i$ is the temperature of the blackbody. 

The fraction of an electron or positron's energy that goes into inverse Compton scattering is then given by:
\begin{eqnarray}
f_{\rm ICS}(E_e) = \frac{\sum_i U_{i, \rm rad} \, \bigg(\frac{\gamma_{i, k}^2}{\gamma_{i, k}^2+\gamma^2}\bigg)}{U_B+\sum_i U_{i, \rm rad} \, \bigg(\frac{\gamma_{i, k}^2}{\gamma_{i, k}^2+\gamma^2}\bigg)}.
\end{eqnarray}

In order to calculate the total spectrum of the resulting electromagnetic cascade, we repeat the above procedure iteratively, evolving each electron/positron until its energy falls below 1 GeV.  In our calculations, we have modelled the radiation fields in a given radio galaxy as a sum of the following blackbodies: the cosmic microwave background (2.7 K), an infrared background from thermal dust emission (50.7 K), and an optical background from starlight (3990 K). While we keep the normalization of the cosmic microwave background fixed to the measured value, we allow the combined infrared and optical intensities to vary within a reasonable range of values in each galaxy. More specifically, we adopt the standard (\ie GALPROP) interstellar radiation field model of the Milky Way as presented in Refs.~\cite{Moskalenko2006,Zhang2006,Porter2005}, and rescale the column depth of the combined optical and infrared components for each individual radio galaxy. We also allow the cascade to continue evolving after exiting the radio galaxy, adopting the Dominguez (2011) model~\cite{Dominguez:2010bv} for the spectrum of the extragalactic background light, and assuming intergalactic magnetic fields to be negligible.

\section{Modeling The Gamma-Ray Emisison From Individual Radio Galaxies}
\label{gammaresults}

In this section, we consider measurements of nearby radio galaxies by Fermi and ground based gamma-ray telescopes and use this information to characterize and constrain the gamma-ray spectra injected from these sources, as well as the energy densities of radiation and magnetic fields that lead to the development of electromagnetic cascades in these systems. 

\begin{figure}[h]
\includegraphics[scale=0.65]{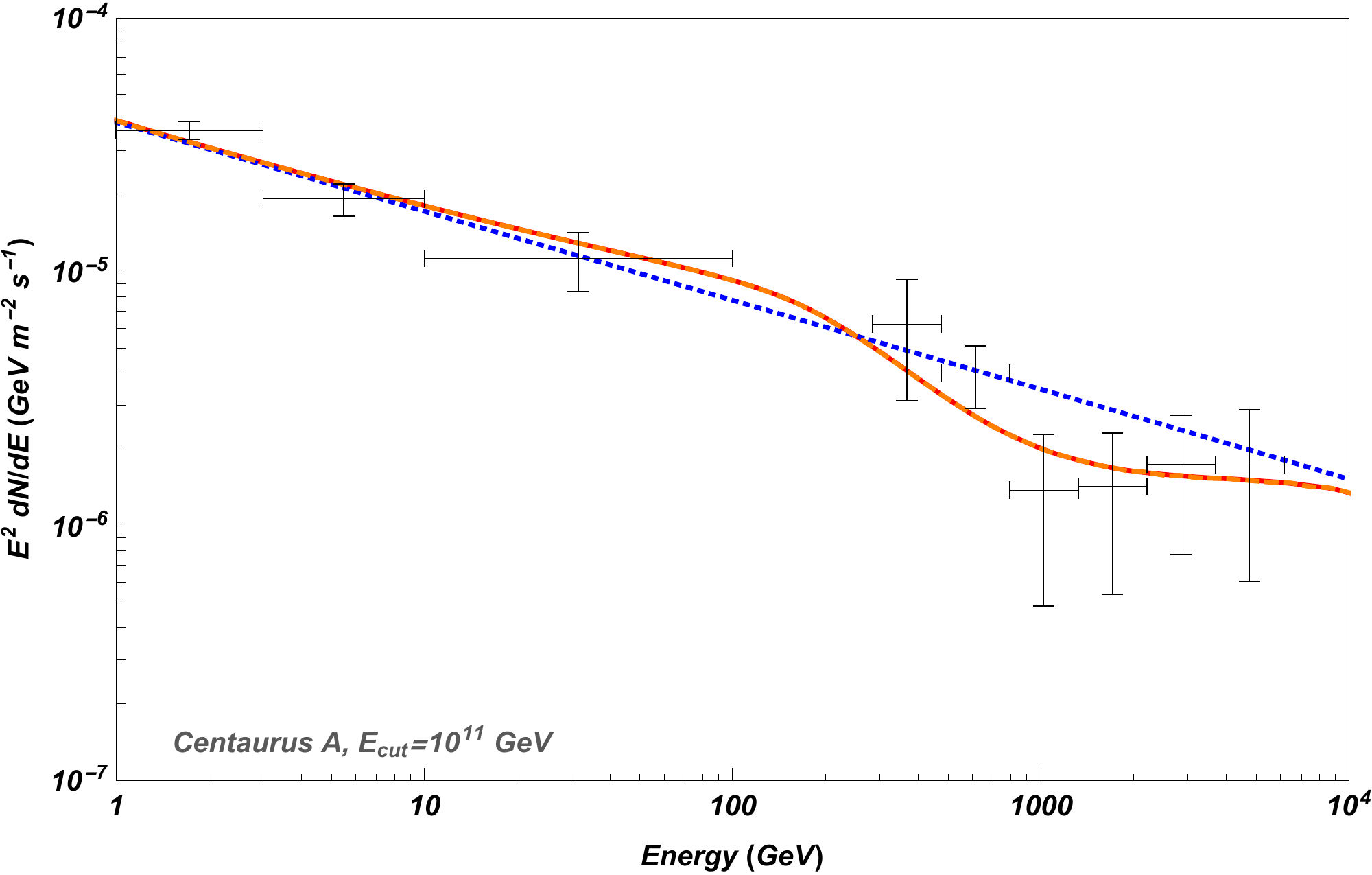} \\
\vspace{0.5cm}
\includegraphics[scale=0.65]
{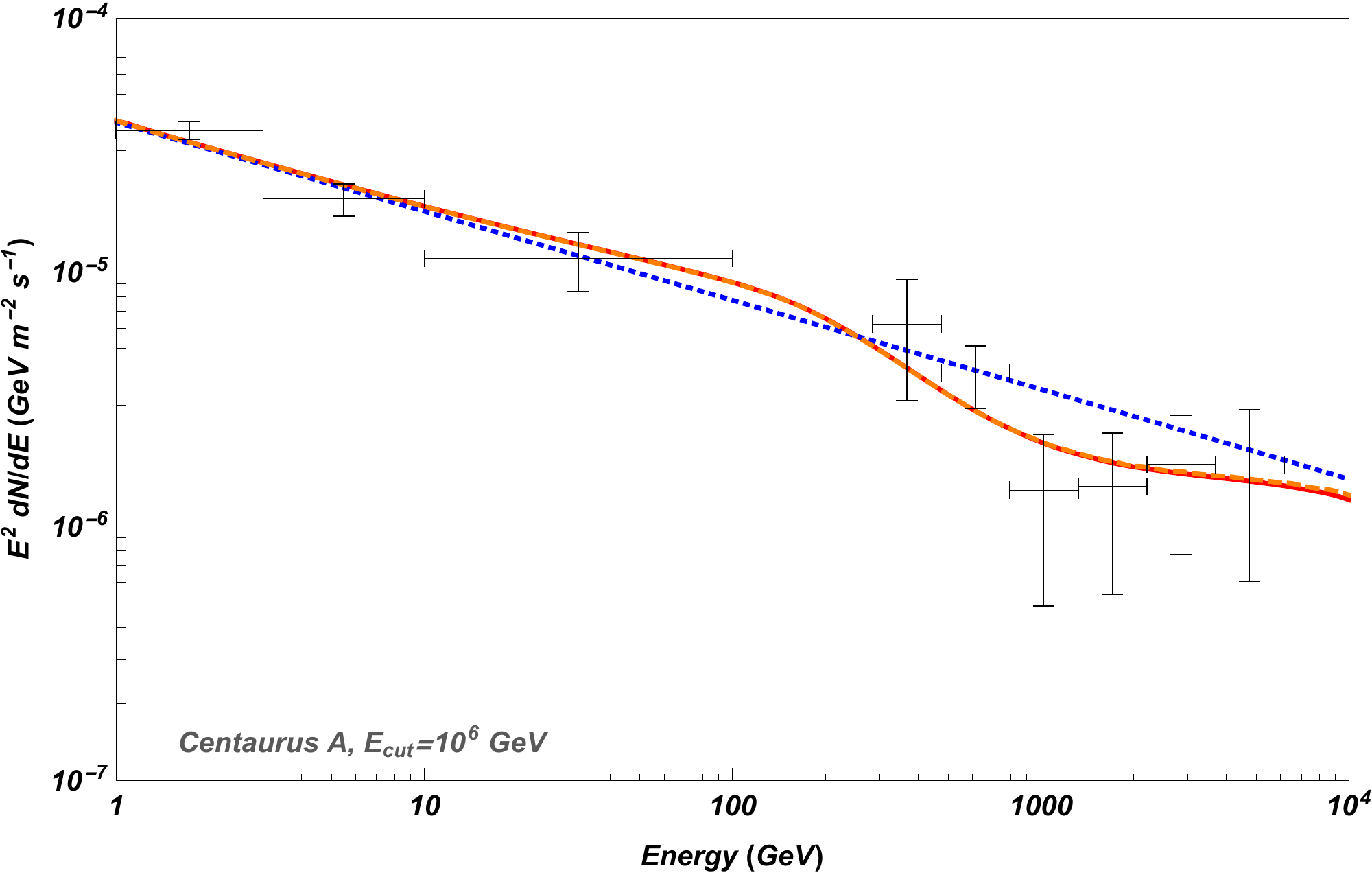}
\caption{The gamma-ray spectrum of the radio galaxy Centaurus A as measured by Fermi (1-100 GeV)~\cite{TheFermi-LAT:2015hja} and HESS ($>$\,200 GeV)~\cite{Aharonian:2009xn}. These results are compared to that predicted by our best-fit spectral index and attenuation/cascade model, and adopting a maximum gamma-ray energy of $E_{\rm cut}=10^{11}$ GeV (top) and $E_{\rm cut}=10^{6}$ GeV (bottom). In each frame, the dotted blue line denotes the injected gamma-ray spectrum, prior to any attenuation or cascade. The dashed orange (solid red) curves describe the spectrum after accounting for interactions within the radio galaxy (and after cosmological propagation).}
\label{specCenA}
\end{figure}

\begin{figure}[h]
\includegraphics[scale=0.65]{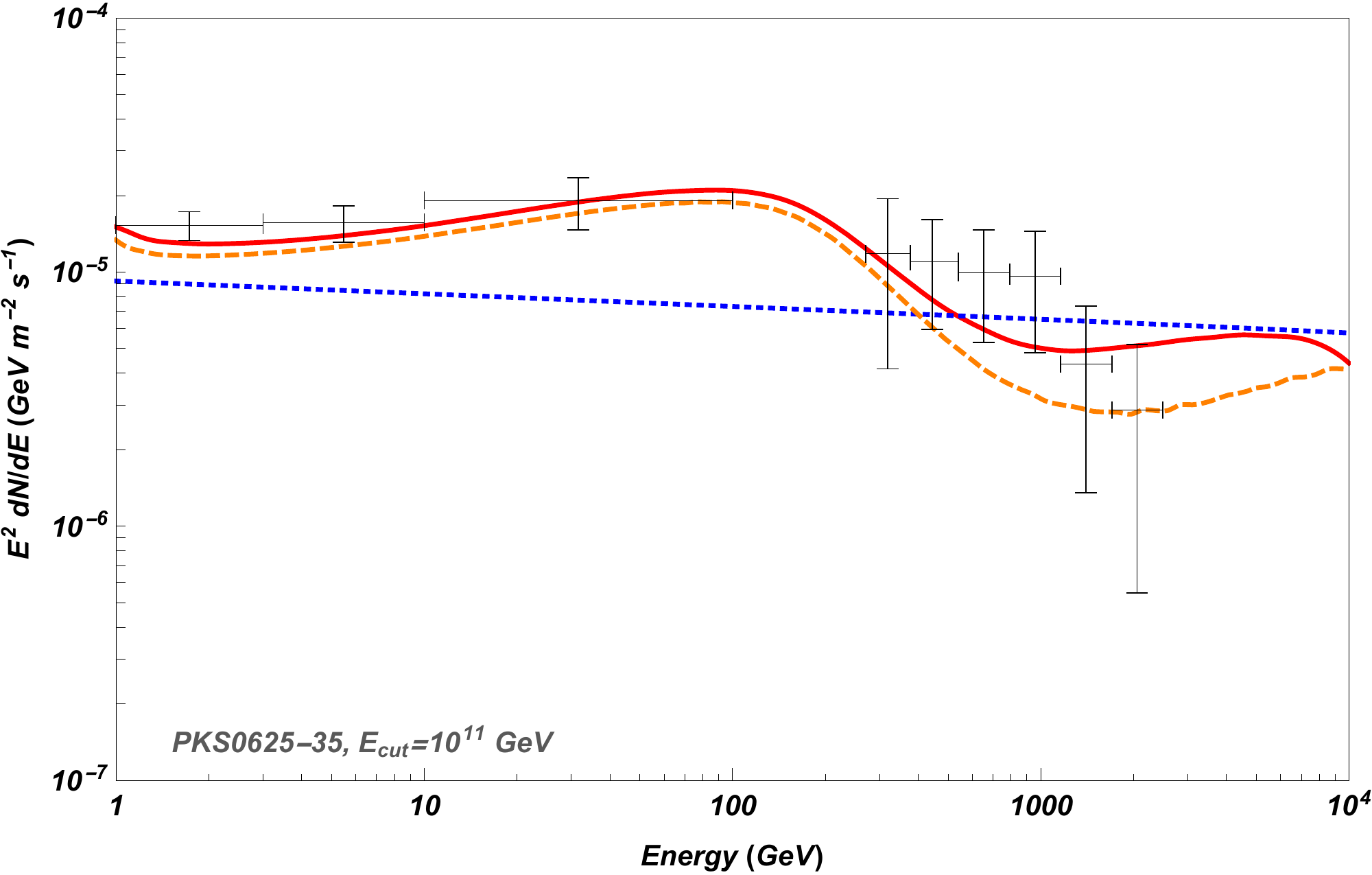} \\
\vspace{0.5cm}
\includegraphics[scale=0.65]{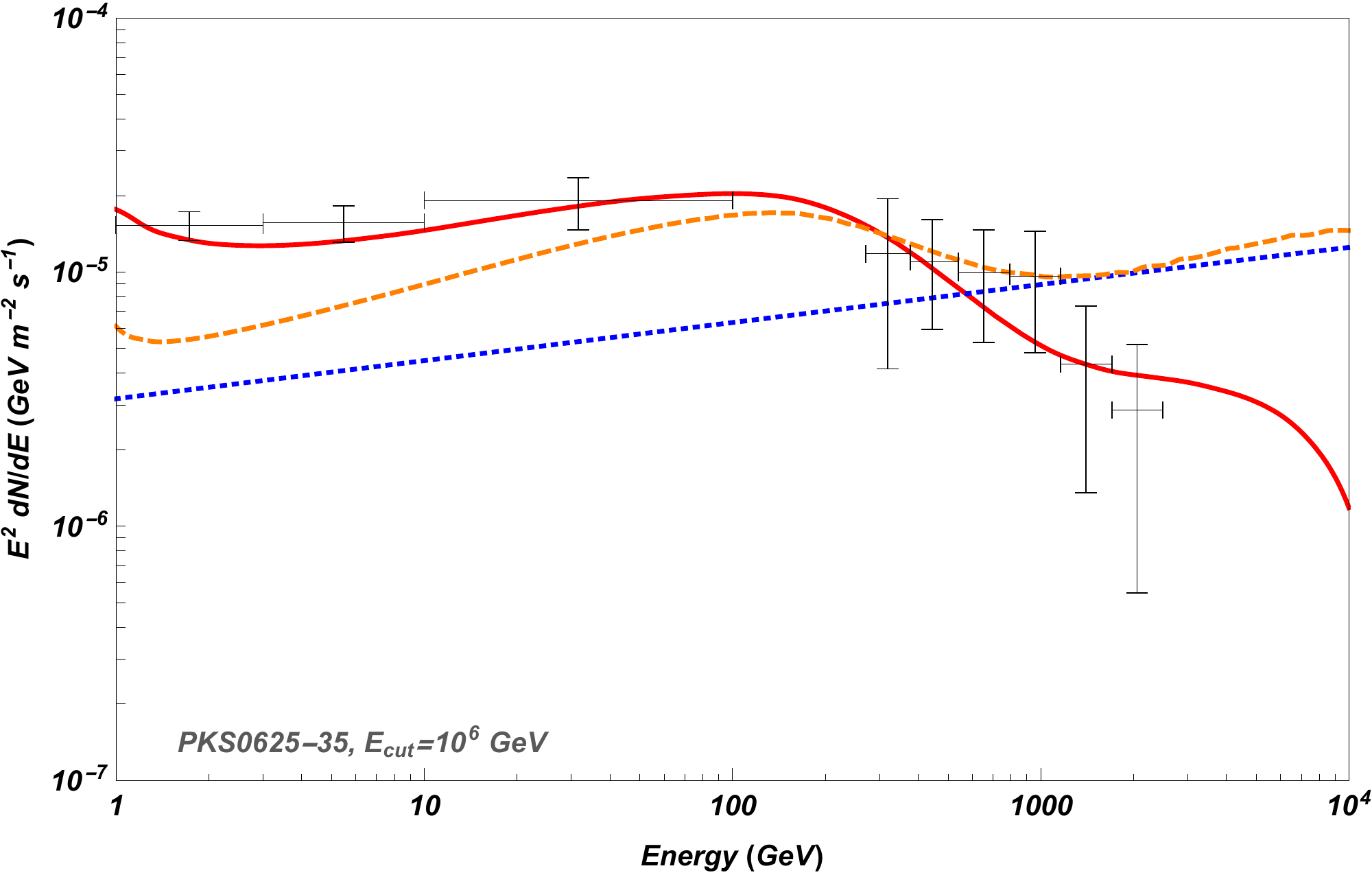}
\caption{As in Fig.~\ref{specCenA}, but for the radio galaxy PKS 0625-35. The measurements shown are those reported by Fermi (1-100 GeV)~\cite{TheFermi-LAT:2015hja} and HESS ($>$\,200 GeV)~\cite{Dyrda:2015hxa}. Again, in each frame, the dotted blue line denotes the injected gamma-ray spectrum, prior to any attenuation or cascade. The dashed orange (solid red) curves describe the spectrum after accounting for interactions within the radio galaxy (and after cosmological propagation).}
\label{specPKS}
\end{figure}

We begin with the radio galaxies Centaurus A and PKS 0625-35, each of which exhibit no discernible variability at GeV or TeV energies.  In Figs.~\ref{specCenA} and~\ref{specPKS}, we present the gamma-ray spectra reported from these sources, including measurements between 1 and 100 GeV by Fermi~\cite{TheFermi-LAT:2015hja}, and at higher energies by HESS~\cite{Aharonian:2009xn,Dyrda:2015hxa}. 

For each source, we compare the measured spectrum to that predicted by our calculations, assuming that the initial injected gamma-ray spectrum is described by a power-law, which is then attenuated by pair production, and supplemented by contributions from electromagnetic cascades. For each galaxy, we vary the following parameters of our model: the intensity and spectral index of the injected gamma-ray emission, the normalization of the combined infrared and optical radiation in the galaxy, and the strength of the magnetic field. For each radio galaxy, we show results assuming that the injected gamma-ray spectrum extends to a maximum energy of either $E_{\rm cut}=10^{11}$ GeV (top frames) or $E_{\rm cut}=10^{6}$ GeV (bottom frames). In each frame, we show the injected gamma-ray spectrum (dotted blue), as well as the spectrum after accounting for interactions within the radio galaxy (dashed orange), and after cosmological propagation (solid red), in each case adopting the best-fit values for the four free parameters in our model.

In the case of Centaurus A, the observed spectrum is indistinguishable from an unbroken power-law, and is best fit by an index of $\gamma=2.38^{+0.11}_{-0.16}$ for $E_{\rm cut}=10^{11}$ GeV (and a nearly identical index for $E_{\rm cut}=10^{6}$ GeV). The spectrum shows no significant evidence of attenuation, or of the presence of a cascade, allowing us to place a modest upper limit on the column depth of radiation in this galaxy equal to $\simeq$\,18 times that of the Milky Way (for the radiation field of the Milky Way, we follow Refs.~\cite{Moskalenko2006,Zhang2006,Porter2005}). Given that little attenuation occurs within the observed energy range in this model, we are not able to meaningfully constrain the magnetic field of this galaxy. We also note that although the spectrum of Centaurus A observed at sub-GeV energies can be explained by synchrotron self-Compton emission~\cite{Falcone:2010fk}, it remains possible that the gamma rays observed at higher energies are generated predominately through hadronic processes.

In the case of PKS 0625-35, the effects of gamma-ray attenuation and the resulting electromagnetic cascade are more evident. The best-fit is found for a radiation column depth that is $\sim$12 (5) times that of the Milky Way for $E_{\rm cut}=10^{11}$ GeV ($10^6$ GeV), although with large uncertainties. A radiation density in the range favored by our fit would not be surprising given the high luminosity of this galaxy. Again, we are not able to significantly constrain the magnetic field energy density, although values in the vicinity of a few $\mu$G provide the best fit. The injected spectral index of this source is somewhat harder than that observed from Centaurus A, $\gamma=2.07 \pm 0.10$ for $E_{\rm cut}=10^{11}$ GeV (and even harder for $E_{\rm cut}=10^{6}$ GeV, for which the best-fit is $\gamma=1.85$). We note that the cascade is more prominent in the case of PKS 0625-35 than in that of Centaurus A in large part because of its harder spectral index, which causes a larger fraction of the total emitted energy to be attenuated and transferred into lower-energy cascade photons.

\begin{figure}[h]
\includegraphics[scale=0.65]{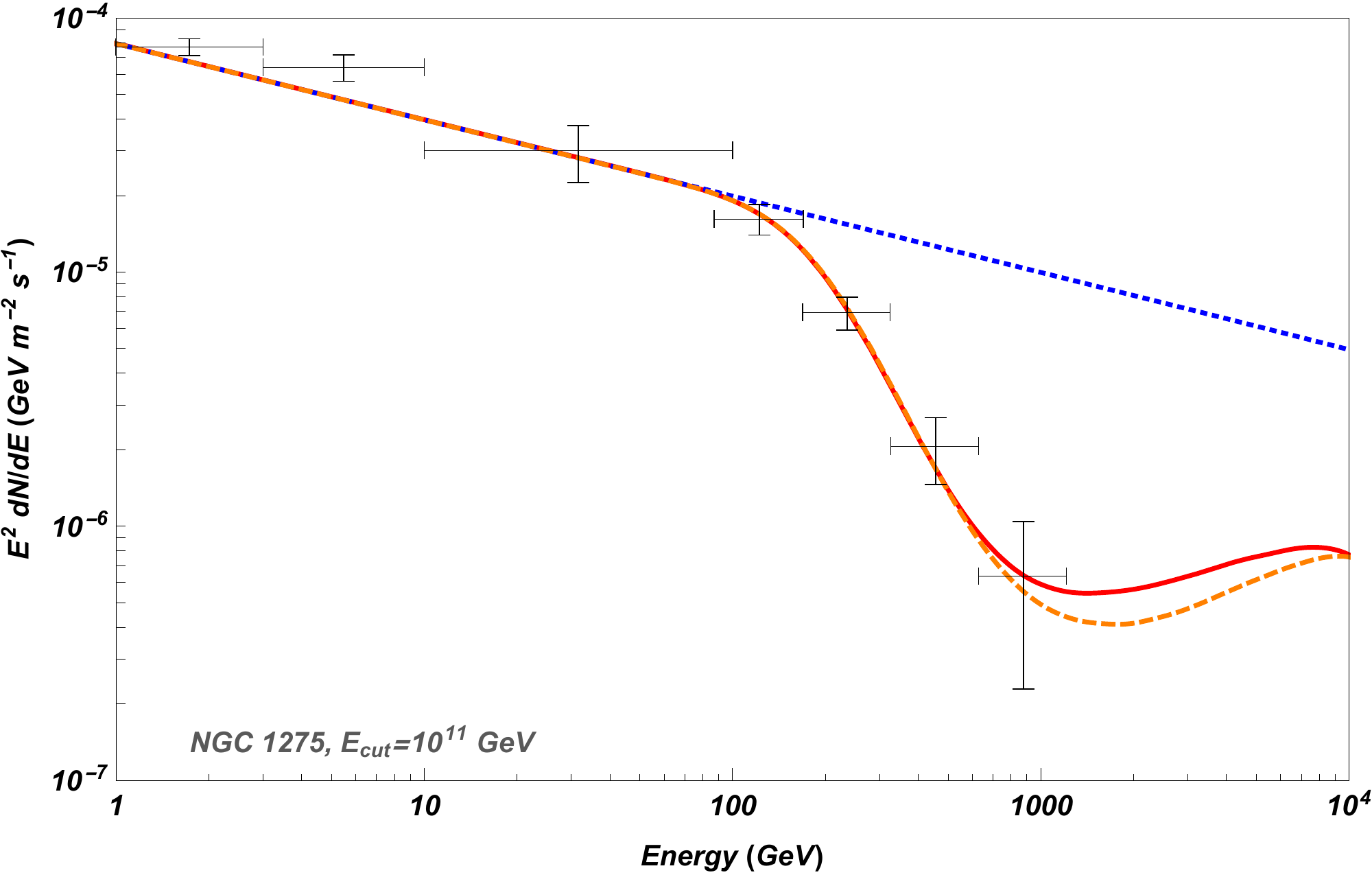} \\
\vspace{0.5cm}
\includegraphics[scale=0.65]{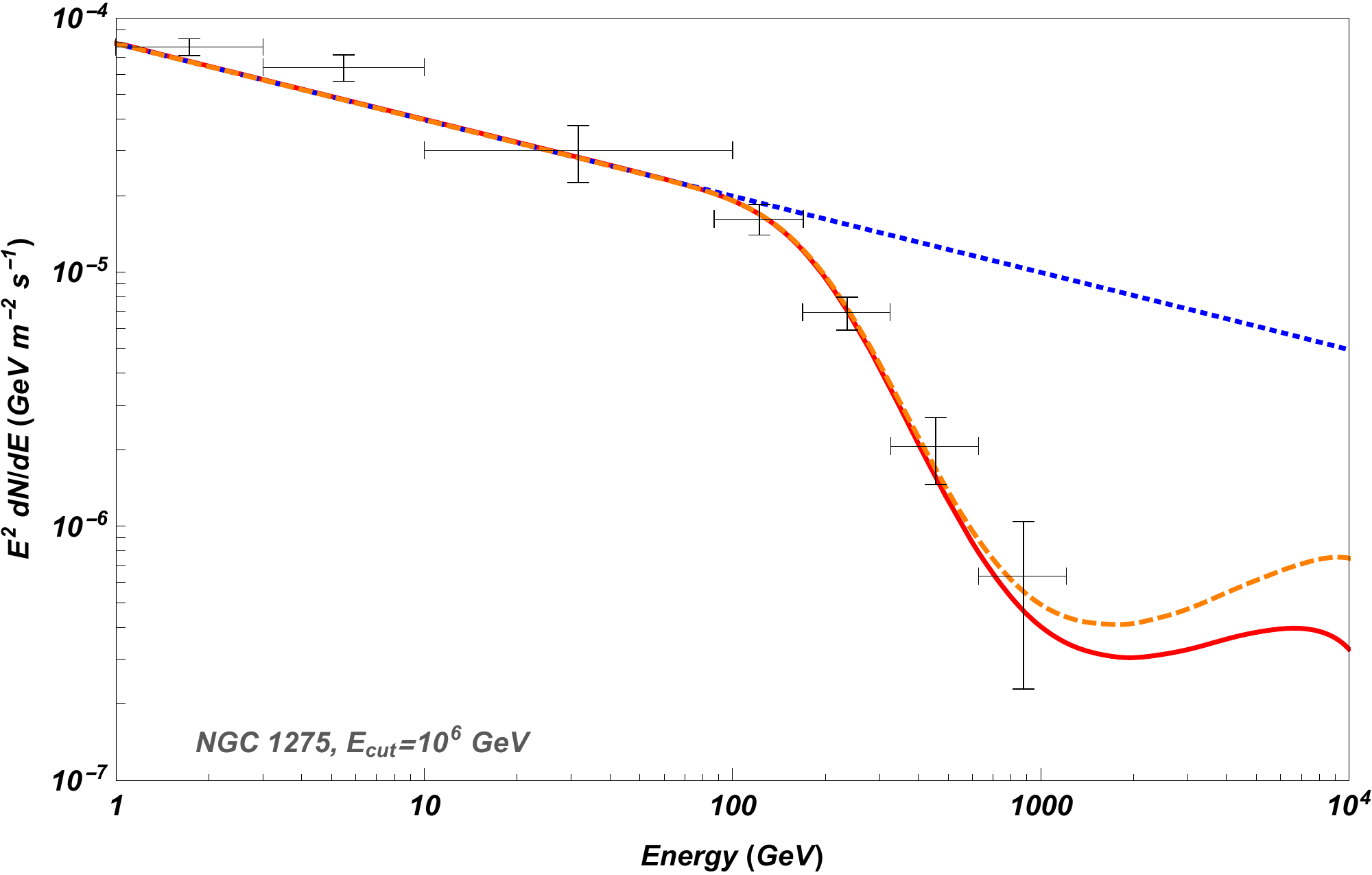}
\caption{As in Figs.~\ref{specCenA}-\ref{specPKS}, but for the radio galaxy NGC 1275. The measurements shown are those reported by Fermi (1-100 GeV)~\cite{TheFermi-LAT:2015hja} and by MAGIC and VERITAS ($\gsim$\,100 GeV)~\cite{Ahnen:2016qkt,2017ATel.9931....1M,2016ATel.9690....1M}. Again, in each frame, the dotted blue line denotes the injected gamma-ray spectrum, prior to any attenuation or cascade. The dashed orange (solid red) curves describe the spectrum after accounting for interactions within the radio galaxy (and after cosmological propagation).}
\label{specNGC}
\end{figure}

Next, we turn our attention the radio galaxy NGC 1275. Unlike either Centaurus A or PKS 0625-35, this source is observed to be variable at GeV energies~\cite{brown2011high}, and to a lesser extent at TeV energies as well~\cite{Ahnen:2016qkt,2017ATel.9931....1M,2016ATel.9690....1M,Galante:2009ie,Aleksic:2013kaa,Conselice2001,Rieger2017,aleksic2014contemporaneous}. From the behavior observed by Fermi, it appears plausible that the emission from NGC 1275 may consist of a combination of variable and steady components, with a ratio of roughly 2:1. With this in mind, in carrying out our fits to NGC 1275 (and as plotted in Fig.~\ref{specNGC}), we reduce the overall normalization of the spectrum reported by Fermi by a factor of three. 

\begin{figure}[h]
\includegraphics[scale=0.65]{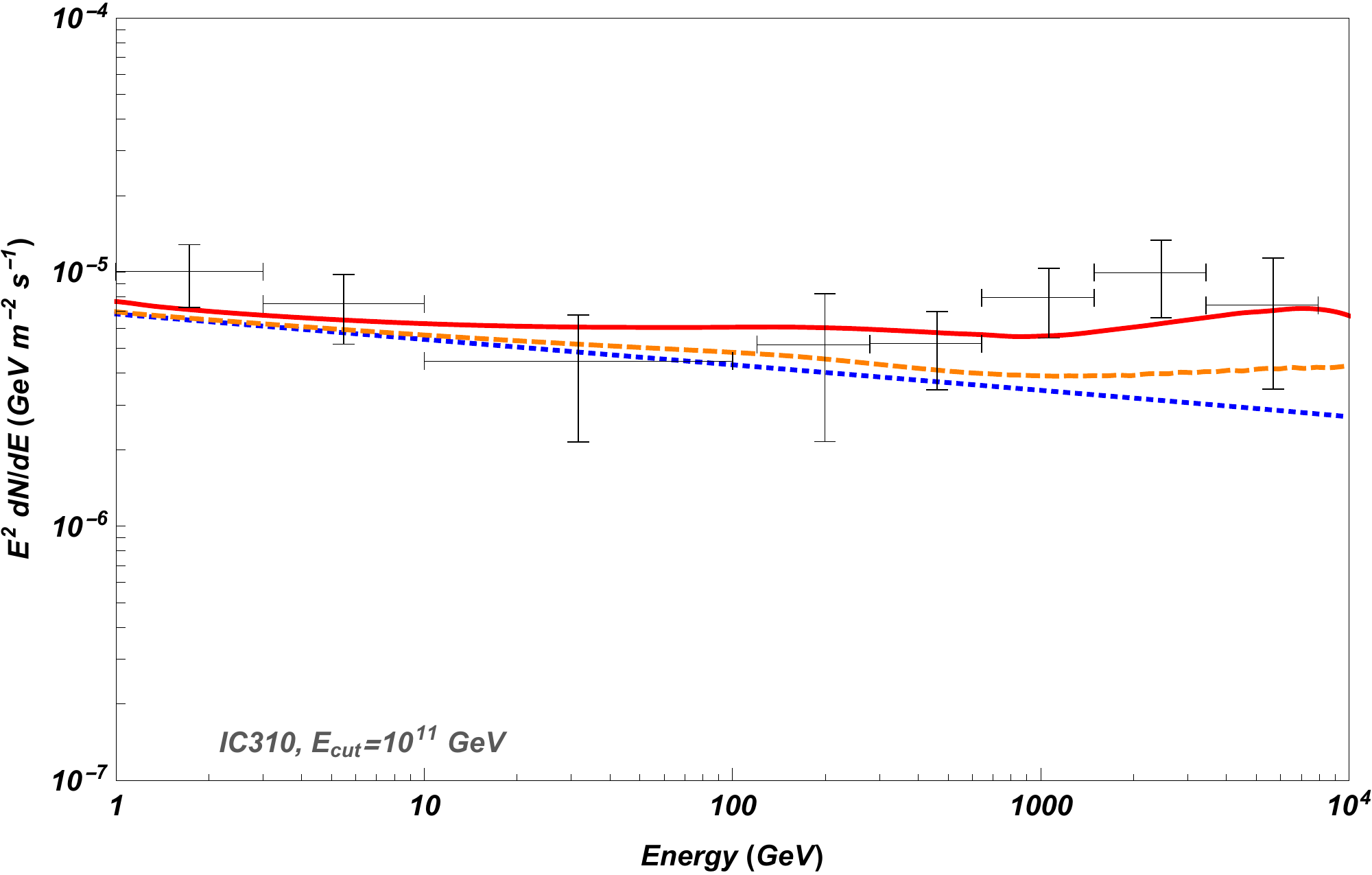} \\
\vspace{0.5cm}
\includegraphics[scale=0.65]{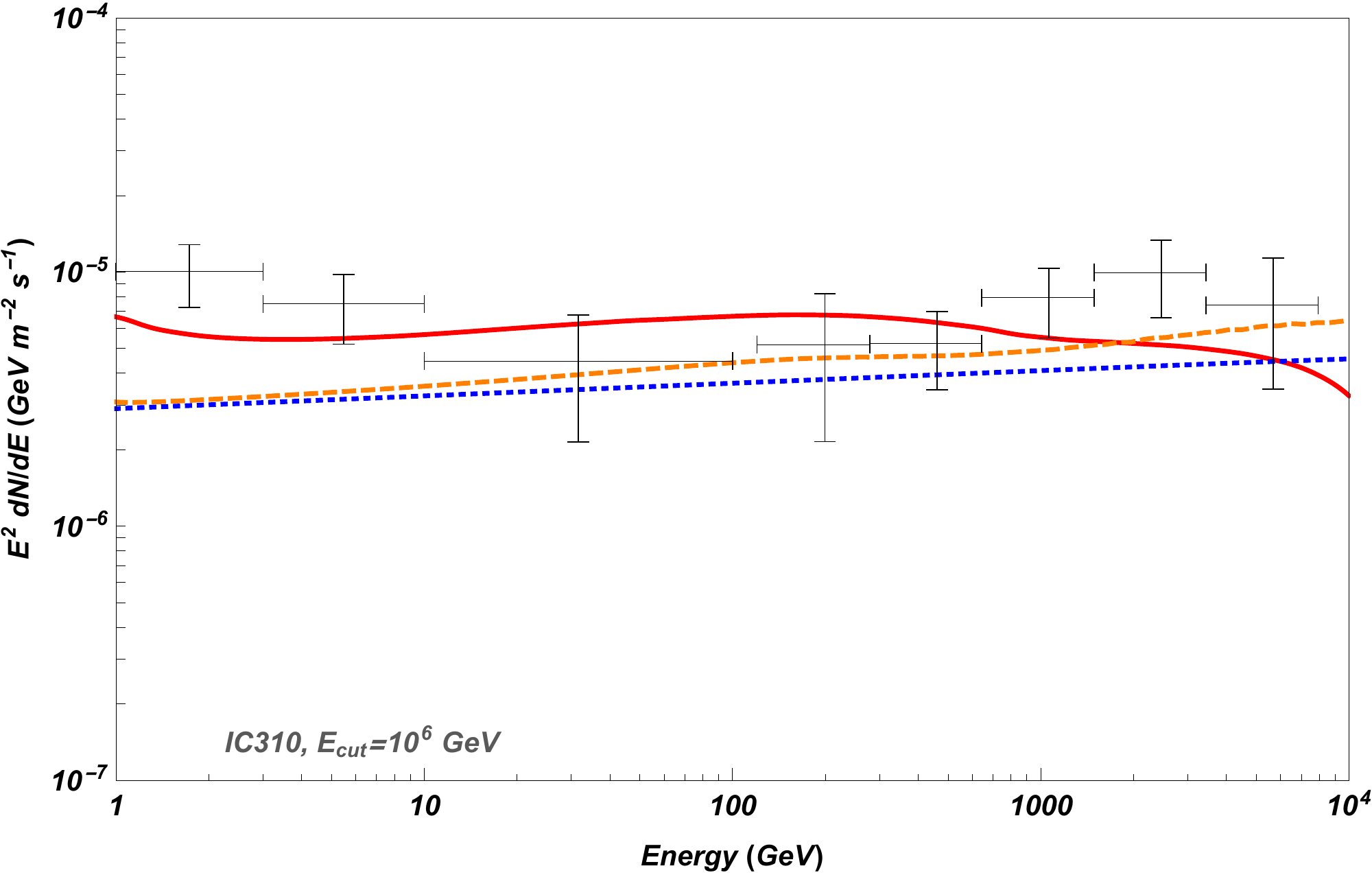}
\caption{As in Fig.~\ref{specCenA}-\ref{specNGC}, but for the radio galaxy IC 310. The measurements shown are those reported by Fermi (1-100 GeV)~\cite{TheFermi-LAT:2015hja} and MAGIC ($\gsim$\,100 GeV)~\cite{Aleksic:2013bya}. Again, in each frame, the dotted blue line denotes the injected gamma-ray spectrum, prior to any attenuation or cascade. The dashed orange (solid red) curves describe the spectrum after accounting for interactions within the radio galaxy (and after cosmological propagation).}
\label{specIC}
\end{figure}

In Fig.~\ref{specNGC}, we show the best-fit spectra that result from this fit. In this case, the effects of attenuation are pronounced, although little evidence of a cascade is present in the spectral shape (due to the relatively soft spectral index of this source). The radiation column depth is found to be $\sim$10-24 times that of the Milky Way in our fit. We speculate that this relatively large value may be, in part, due to the presence of the high-velocity system located along the line-of-sight~\cite{yu2015high,gillmon2004}. Again, we are not able to significantly constrain the magnetic field energy density. The spectral index of this source is $\gamma=2.32 \pm 0.08$ for $E_{\rm cut}=10^{11}$ GeV (and $\gamma=2.33 \pm 0.10$ for $E_{\rm cut}=10^{6}$ GeV).

And lastly, in Fig.~\ref{specIC} we show the results of our fit to the spectrum of the radio galaxy IC 310. For a choice of $E_{\rm cut}=10^{11}$ GeV, we find that the fit prefers an injected spectral index of $\gamma = 2.09^{+0.08}_{-0.15}$. Although the best-fit was found for a radiation column density near that of the Milky Way and a negligible magnetic field, we were not able to meaningfully constrain these quantities. Similar to PKS 0625-35, we find that the required spectral index becomes harder if we reduce the value of the $E_{\rm cut}$. For a choice of $E_{\rm cut}=10^{6}$ GeV, for example, we find a best-fit of $\gamma=1.97$.


At this point, we can use the results presented in this section to make a few general comments. If the extraterrestrial neutrino flux observed by IceCube originates from proton-proton interactions, the injected fluxes of gamma rays and neutrinos are predicted to feature a common spectral index. Given that IceCube has measured a spectral index in the approximate range of $\gamma_{\nu} \sim 2.0-2.5$~\cite{Aartsen:2016xlq,Aartsen:2015rwa,Aartsen:2015knd}, we favor radio galaxy models with values of $\gamma$ in approximately this range. Interestingly, we find that this range is in good agreement with the results presented in this paper. For $E_{\rm cut} =10^{11}$ GeV, all four radio galaxies considered feature best-fit indices within this preferred range ($\gamma=2.38$ for Centaurus A, 2.07 for PKS 0625-35, 2.32 for NGC 1275, and 2.09 IC 310). In contrast, for a lower cutoff of $E_{\rm cut} =10^6$ GeV, we find that PKS 0625-35 and IC 310 each favor significantly harder spectral indices.\footnote{The spectral indices favored by our fit are generally harder for lower values of $E_{\rm cut}$. This is because photons in the highest range of energies deposit most of the energy into cascades which peak in the VHE band, whereas the cascades from lower energy photons peak in the $\sim$1-100 GeV range.} This fact provides support for scenarios in which IceCube's spectrum extends to energies well above the PeV scale, and for the possibility that this class of astrophysical objects is responsible for the production of the ultra-high energy cosmic rays.

\newpage

\section{Pair Halos and the Contribution From Radio Galaxies to the Isotropic Gamma-Ray Backgorund}

In the previous section, we found that a significant fraction of the emission observed from radio galaxies in the GeV-TeV range may be the result of electromagnetic cascades produced through the pair production of VHE photons with radiation backgrounds. This is especially true in the cases of PKS 0625-35 and and IC 310, which each exhibit somewhat hard spectral indices. 

When a VHE photon undergoes pair production and initiates an electromagnetic cascade, this has the effects of broadening the angular width of the emission observed from the source. In particular, the cascade emission from a source at a distance $D$ will approach us from the following angle relative to that of the source:
\begin{equation}
\sin \theta_{\rm Ext} = \frac{d}{D} \, \sin \delta,
\end{equation}
where the VHE photon initiates the cascade after propagating a distance, $d$, and $\delta$ is the angle between the propagation direction of the cascade photon and that of the initial VHE photon. This angle can be approximate by the following~\cite{Neronov:2009gh,Tashiro:2013bxa,Chen:2014rsa}:
\begin{eqnarray}
\sin \delta \simeq  \frac{D_e}{r_L} \sim 3\times 10^{-3} \, \bigg(\frac{B}{10^{-15} \, {\rm G}}\bigg) \, \bigg(\frac{10 \, {\rm TeV}}{E_e}\bigg)^2,
\end{eqnarray}
which for $\sin \delta \ll 1$ reduces to the following:
\begin{eqnarray}
\delta \simeq  \frac{D_e}{r_L} \sim 0.2^{\circ} \, \bigg(\frac{B}{10^{-15} \, {\rm G}}\bigg) \, \bigg(\frac{10 \, {\rm TeV}}{E_e}\bigg)^2.
\end{eqnarray}
In these expressions, $D_e$ is the length scale over which an electron of energy $E_e$ loses its energy via inverse Compton scattering and $r_L$ the Larmor radius of the same electron in a magnetic field of strength $B$. 

As a concrete example, consider a source located at a distance of 100 Mpc. If the strength of the intergalactic magnetic field is $B \sim 10^{-16}$ G or less, essentially the entire energy range that is efficiently attenuated will feature $\delta \lsim 0.1^{\circ}$, and thus the morphology of the resulting cascade emission will be indistinguishable from that of the original point source. If the intergalactic magnetic field is stronger, however, a significant fraction of the cascade could be observably deflected.

\begin{figure}[h]
\includegraphics[scale=0.65]{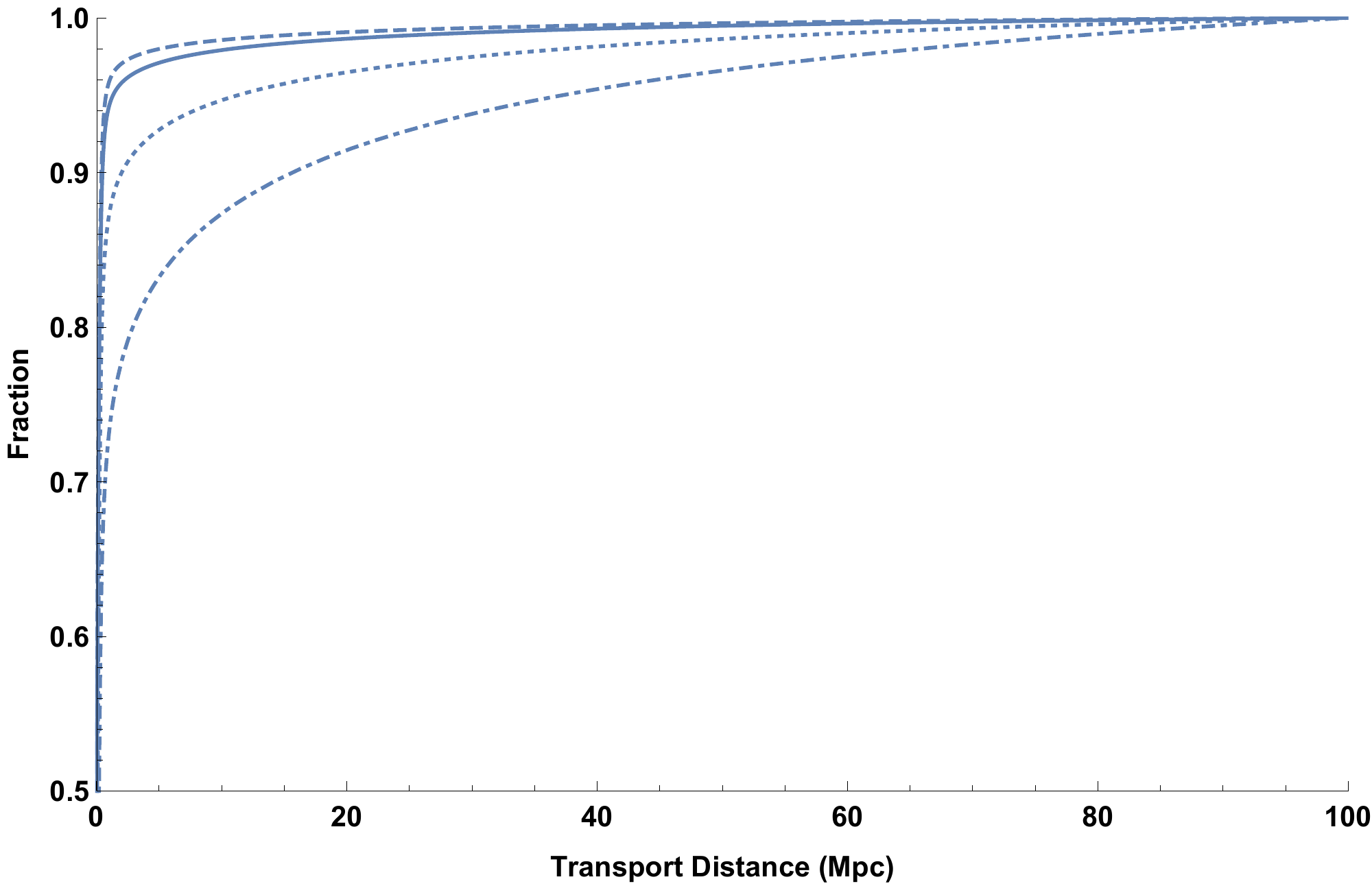}
\caption{The fraction of the total cascade emission that is generated as a function of distance, for a source that is located at a distance of 100 Mpc. Results are shown for spectral indices of 1.8 (dashed), 2.0 (solid), 2.2 (dotted) and 2.4 (dot-dashed). For a spectral index of $\gamma=2.2$, approximately 90\% of the total cascade energy is deposited within the first 2 Mpc around the source.}
\label{evolution}
\end{figure}

For an approximately maximal value of $B \sim 10^{-10}$ G, for example, the resulting inverse Compton emission generated in the cascade will be approximately isotropic for electrons less energetic than $E_e \lsim 200$ TeV. As a consequence of these deflections, we predict that approximately half of the total cascade emission will not be directed toward the source in this case, but will instead contribute to the IGRB (for the case of $\gamma=2.2$ and $E_{\rm cut}=10^{11}$ GeV). Furthermore, those electrons which are either more modestly deflected, or that are deflected within a few Mpc of the source, can lead to the appearance of a ``pair halo'' of gamma-ray emission, extending around the source by $\sim0.1-1^{\circ}$. In this case, the main contribution to the pair halo results from electrons with energies of order $\sim$100 TeV and that are produced at a distance of approximately $\sim$0.3-3 Mpc from the source.\footnote{In Fig.~\ref{evolution}, we plot the fraction of the total cascade emission that is generated as a function of distance (for a source located at 100 Mpc). For a spectral index of $\gamma=2.2$, approximately 90\% of the total cascade energy is deposited within 2 Mpc of the source.} For this choice of distance and magnetic field strength, we expect that approximately $\sim$5-10\% of the total cascade emission will contribute to the formation of a pair halo. 

For intermediate values of the magnetic field strength ($B \sim 10^{-14}-10^{-11}$ G), we expect most of the cascade emission to be indistinguishable from the original point source, although non-negligible pair halos can be generated over much of this parameter space.

\section{Connections with Neutrinos and Ultra-High Energy Cosmic Rays}

Active galaxies are among the most promising sources for the acceleration of high-energy cosmic rays. And although many models have been discussed in recent years (see, for example, Ref.~\cite{Murase:2015ndr,Tjus:2014dna,Kimura:2014jba,Caprioli:2015zka}), it is not difficult to imagine scenarios in which these cosmic rays interact with gas to generate both the gamma-rays observed from these sources, as well as the diffuse neutrino flux observed by IceCube. In this section, we consider the results of the previous sections within the context of a simple model in which many of the cosmic rays accelerated by active galaxies escape the acceleration region and then proceed to diffuse through the surrounding galactic (or perhaps cluster) environment, where a fraction of them scatter with gas to generate charged and neutral pions. Throughout this discussion, we will remain agnostic as to where and by what means the cosmic-ray acceleration is taking place within active galaxies, only assuming that after being accelerated, these particles undergo diffusion through the surrounding medium.

Over a time, $t$, a typical cosmic ray will travel a net distance given by $d_{\rm dif} \sim 2\sqrt{D(E_p) t}$, where $D(E_p)$ is the energy dependent diffusion coefficient. Adopting a Kolmogorov spectrum of magnetic inhomogeneities, the diffusion coefficient takes the following form:
\begin{eqnarray}
\label{dif}
D(E_p) = \frac{1}{3} c \, l_c \bigg(\frac{r_L}{l_c}\bigg)^{1/3}  \approx  \, 1.5 \times 10^{30} \, {\rm cm}^2/{\rm s} \, \bigg(\frac{E_p}{{\rm PeV}}\bigg)^{1/3} \bigg(\frac{l_c}{{\rm kpc}}\bigg)^{2/3} \bigg(\frac{ \mu{\rm G}}{B}\bigg)^{1/3}, 
\end{eqnarray}
where $r_L$ is the Larmor radius of the propagating cosmic ray and $l_c$ is the coherence length of the magnetic field. Our representative choice of $l_c^2/B \simeq {\rm kpc}^2/\mu{\rm G}$ should be considered a reasonable estimate, as this leads to a diffusion coefficient that is consistent with that derived from measurements of GeV-TeV cosmic rays in the Milky Way~\cite{Simet:2009ne,Trotta:2010mx}. Diffusion is expected to continue up to energies of $E_p \sim {\rm EeV} \times (l_c/{\rm kpc}) (B/\mu{\rm G})$, above which free-streaming is expected to dominate, allowing cosmic rays to more rapidly escape from their parent galaxies.

From the diffusion coefficient given in Eq.~\ref{dif}, we can estimate that a typical cosmic-ray proton will will remain confined within the volume of its parent galaxy for the following time:
\begin{eqnarray}
t_{\rm esc} \sim \frac{d^2_{\rm dif}}{4 D(E_p)}  \sim  1.6 \times 10^{14} \, {\rm s} \times \bigg(\frac{d_{\rm dif}}{20\,{\rm kpc}}\bigg)^2 \bigg(\frac{{\rm PeV}}{E_p}\bigg)^{1/3}  \bigg(\frac{{\rm kpc}}{l_c}\bigg)^{2/3} \bigg(\frac{B}{\mu{\rm G}}\bigg)^{1/3},
\end{eqnarray}
during which, the probability of scattering with gas is given by $P(E_p) = 1-e^{-\tau_{pp}(E_p)}$, where the optical depth is given by:
\begin{eqnarray}
\label{tau}
\tau_{pp}(E_p) &\sim& \sigma_{pp}(E_p) \, c \, t_{\rm esc} \, n_{\rm gas} \\
& \sim & 0.04 \times \bigg(\frac{n_{\rm gas}}{0.05 \, {\rm cm}^{-3}}\bigg) \bigg(\frac{d_{\rm dif}}{20\,{\rm kpc}}\bigg)^2 \bigg(\frac{{\rm PeV}}{E_p}\bigg)^{1/3}  \bigg(\frac{{\rm kpc}}{l_c}\bigg)^{2/3} \bigg(\frac{B}{\mu{\rm G}}\bigg)^{1/3}. \nonumber
\end{eqnarray}
In this expression, $n_{\rm gas}$ represents the average number density of targets within the diffusion region. The value of the gas density, $0.05$~cm$^{-3}$, adopted in the above expression is appropriate for systems such as NGC 1275~\cite{Churazov:2003hr}, but likely overestimates the optical depth for most other radio galaxies~\cite{1986ApJ...310..637T,Kraft:2003gp,Allen:2006mh,Fukazawa:2005nx,Taylor:2001jw,Kim:1997dv,Owen:2000vi}. The reader should remain open-minded to the possibility that the high-energy neutrino flux observed by IceCube may arise from a subset of radio galaxies (or galaxy clusters containing active galaxies), weighted by the relative densities of gas, averaged across their diffusion regions. In any case, the value of the optical depth does not directly enter any of the calculations or figures presented in this study.

Adopting a power-law form for the spectrum of protons accelerated by the active galaxy, $dN_p/dE_P = A_p E_p^{-\Gamma_p}$, the resulting spectra of gamma rays and neutrinos are given as follows:
\begin{eqnarray}
\label{gammaspec}
\frac{dN_{\gamma}}{dE_{\gamma}} &\approx& A_{\gamma} \times  \bigg(\frac{E_{\gamma}}{{\rm GeV}}\bigg)^{-\frac{4}{3}\Gamma_p + \frac{1}{3}},
\end{eqnarray}
and
\begin{eqnarray}
\label{neutrinospec}
\frac{dN_{\nu}}{dE_{\nu}} &\approx& \frac{3}{2} A_{\gamma}  \times  \bigg(\frac{E_{\nu}}{{\rm GeV}}\bigg)^{-\frac{4}{3}\Gamma_p + \frac{1}{3}},
\end{eqnarray}
where the normalization constant $A_{\gamma}$ is given by the relationship $\int A_p E_p^{-\Gamma_p+1} \tau_{pp}(E_p) dE_p = 3 \int A_{\gamma} E_{\gamma}^{-(4/3)\Gamma_p+(4/3)}  dE_{\gamma}$. These power-law spectra are expected to extend up to energies of approximately $E_{\gamma} \sim 10^8 \, {\rm GeV} \times (l_c/{\rm kpc}) (B/\mu{\rm G})$ and $E_{\nu} \sim 0.5 \times 10^8 \, {\rm GeV} \times (l_c/{\rm kpc}) (B/\mu{\rm G})$, above which cosmic-ray free-streaming is expected to lead to a steepening of the spectral index~\cite{Aloisio:2004jda}. In this calculation, we have made use of the fact that the average number of pions produced in a proton-proton collision scales as $N_{\pi} \propto E_p^{1/4}$ over the relevant range of energies, and the average fraction of energy that is carried by a given pion scales as $\langle E_{\pi} \rangle /E_p \propto  E_p^{-1/4}$~\cite{1994A&A...286..983M,Tjus:2014dna}.

In order to accommodate the range of injected gamma-ray spectral indices that were found to be favored in Sec.~\ref{gammaresults} ($\gamma \simeq 2.0-2.4$), we find from Eq.~\ref{gammaspec} that protons must be injected with a spectral index of $\Gamma_p = 0.75 \, \gamma +0.25 \simeq 1.75-2.05$. This range of spectral indices, as well as the measured redshift distribution of radio galaxies, is in good agreement with that required to accommodate the observed ultra-high energy cosmic ray spectrum~\cite{Taylor:2015rla,Taylor:2013gga,Taylor:2011ta}. 

For our choice of parameters, electrons more energetic than a few hundred TeV lose their energy primarily through synchrotron emission. For a radio galaxy with an injected gamma-ray spectral index of $\gamma=-2$ extending up to $10^{11}$ GeV, and normalized to the gamma-ray flux at 1 GeV from Centaurus A, for example, we estimate that this leads to a synchrotron spectrum of $\sim 10^{-6} \, (E/{\rm GeV})^{-2.5}$ GeV$^{-1}$m$^{-2}$s$^{-1}$ emitted over a range of critical frequencies from $\sim$10 keV to $\sim$10 GeV. This constitutes only a small fraction of the total gamma-ray emission (see Fig.~\ref{specCenA}) and X-ray emission~\cite{Beckmann:2011rh} from such a source.

Given the relatively modest mean free paths of the highest energy cosmic rays, one could expect the ultra-high energy cosmic ray spectrum to be dominated by the nearest radio galaxies. Within this context, it is particularly interesting that the Auger Collaboration has reported a modest excess of events above 55 EeV from directions within $\sim 20^{\circ}$ of Cen A (see Fig.~9 of Ref.~\cite{Abreu:2010ab}, and also Refs.~\cite{Liu:2012sq,Biermann:2011wf,Yuksel:2012ee,Kim:2012rp,Farrar:2012gm,Sushchov:2012ri,Keivani:2014kua}). Although this signal constitutes only a $\sim2 \sigma$ excess, it is suggestive within the context of the scenario considered here.

To estimate the degree to which a given ultra-high energy cosmic ray will be deflected, consider a particle of charge $Z$, which traverses a distance $l$ though a uniform magnetic field. Such a particle will be deflected by an angle of $\theta_0=l/r_L$, where $r_L$ is the Larmor radius of the particle. Over a large distance, $D$, such a particle will pass through many such regions, each with a size comparable to the coherence length of the magnetic field, $l_c$. This leads to a total deflection that is given by $\theta \approx \theta_0 \, \sqrt{D/\l_{\rm coh}} \approx \sqrt{D \, \l_{\rm coh}}/r_L$. In the case of Centaurus A ($D\approx4$ Mpc), we estimate that extragalactic magnetic fields will deflect ultra-high energy cosmic rays by the following angle:
\begin{eqnarray}
\theta_{\rm EG} \approx 1.1^{\circ} \times \bigg(\frac{10^{20}\, {\rm eV}}{E}\bigg)\, \bigg(\frac{D}{4 \, {\rm Mpc}}\bigg)^{0.5} \,  \bigg(\frac{l_{\rm c}}{{\rm Mpc}}\bigg)^{0.5} \,  \bigg(\frac{B}{10^{-10} \, {\rm G}}\bigg) \, \bigg(\frac{Z}{10}\bigg), 
\end{eqnarray}
where $E$ is the energy of the cosmic ray, and $B$ is the strength of the extragalactic magnetic field. From this estimate, we conclude that the $\sim\,20^{\circ}$ deflections suggested by the Auger data are unlikely to result from extragalactic magnetic fields. On the other hand, the magnetic field of the Milky Way is expected to deflect cosmic rays to a greater degree:
\begin{eqnarray}
\theta_{\rm MW} \approx 17^{\circ} \times \bigg(\frac{10^{20}\, {\rm eV}}{E}\bigg)\, \bigg(\frac{D}{10 \, {\rm kpc}}\bigg)^{0.5} \,  \bigg(\frac{l_{\rm c}}{{\rm kpc}}\bigg)^{0.5} \,  \bigg(\frac{B}{10^{-6} \, {\rm G}}\bigg) \, \bigg(\frac{Z}{10}\bigg). 
\end{eqnarray}
This result is in good agreement with the angular extent of the excess above 55 EeV tentatively observed in the direction around Centaurus A by the Auger Collaboration.

\section{The Predicted Spectrum of High-Energy Neutrinos}

Assuming that the observed gamma-ray emission from a given radio galaxy is generated through the interactions of cosmic-ray protons with gas, we can directly relate the injected gamma-ray spectrum to the predicted spectrum of high-energy neutrinos. In particular, it follows from Eqs.~\ref{gammaspec} and~\ref{neutrinospec} that the gamma ray and neutrino spectra will share a common spectral index, and will exhibit overall relative fluxes given by $F_{\nu}/F_{\gamma} \approx 2 \times (3/4) = 3/2$, where the factors of 2 and 3/4 correspond to the ratio of charged-to-neutral pions produced and to the fact that three of the four $\pi^{\pm}$ decay products are neutrinos. 

\begin{figure}[h]
\includegraphics[scale=0.65]{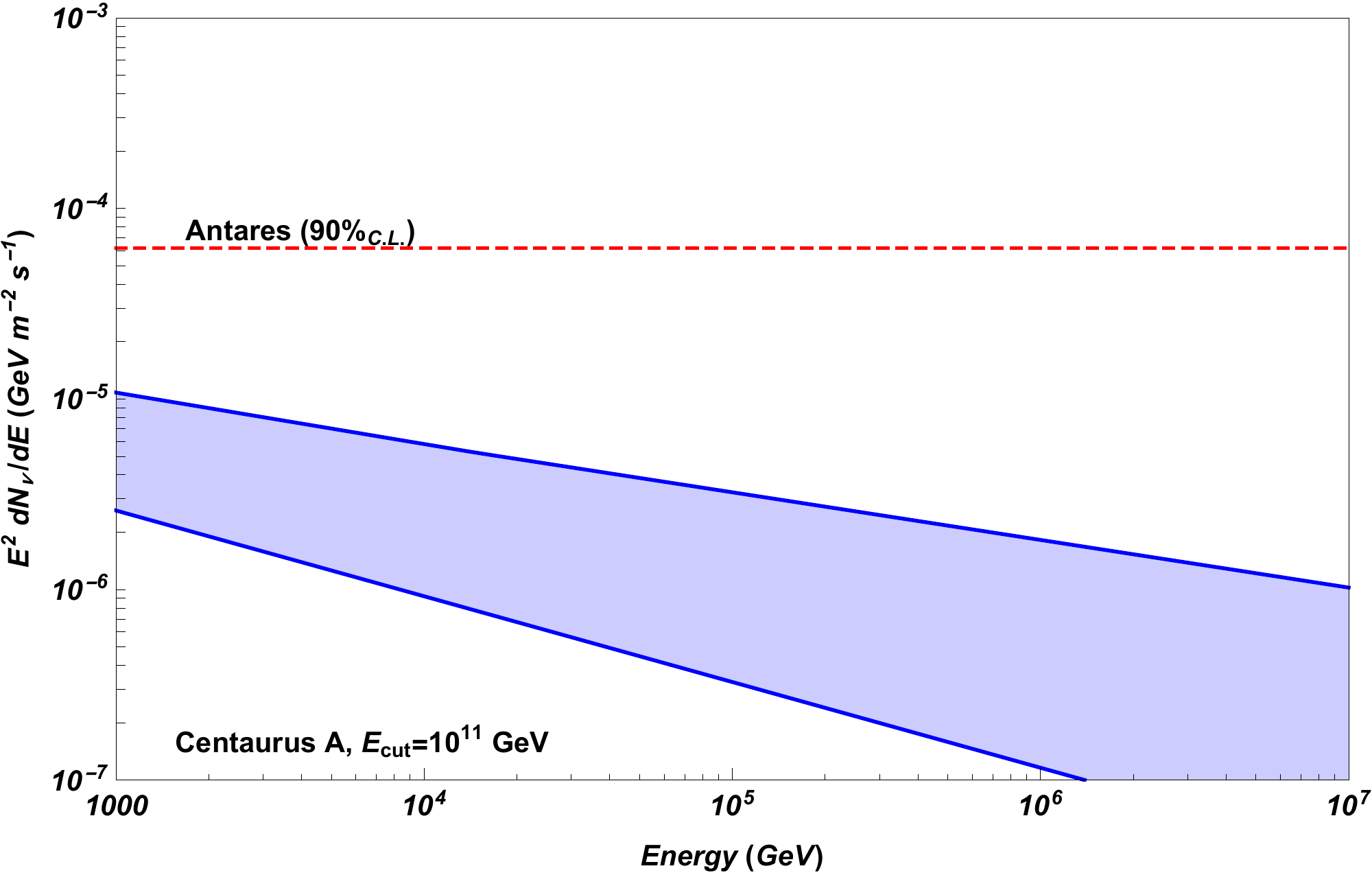} \\
\vspace{0.5cm}
\includegraphics[scale=0.65]{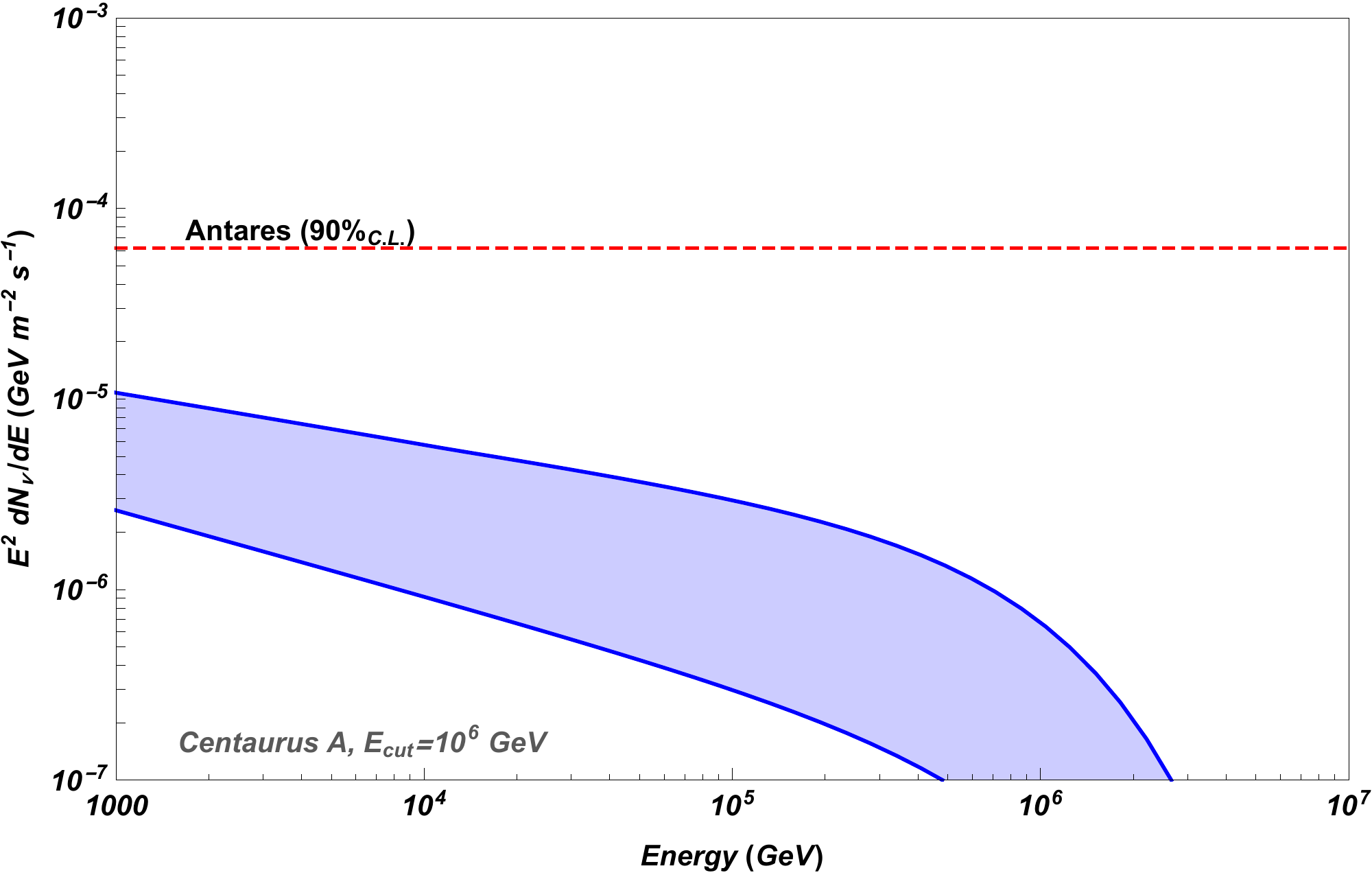}
\caption{The neutrino spectrum predicted from the radio galaxy Centaurus A, adopting a maximum gamma-ray energy of $E_{\rm cut}=10^{11}$ GeV (top) and $E_{\rm cut}=10^{6}$ GeV (bottom). In each frame, the band shown reflects the range predicted after marginalizing over the spectral index and normalization in the fit to the observed gamma-ray spectrum. Also shown is the (90\% confidence level) upper limit from this source as reported by the ANTARES Collaboration~\cite{Albert:2017ohr}.}
\label{neutrinospecCenA}
\end{figure}

\begin{figure}[h]
\includegraphics[scale=0.65]{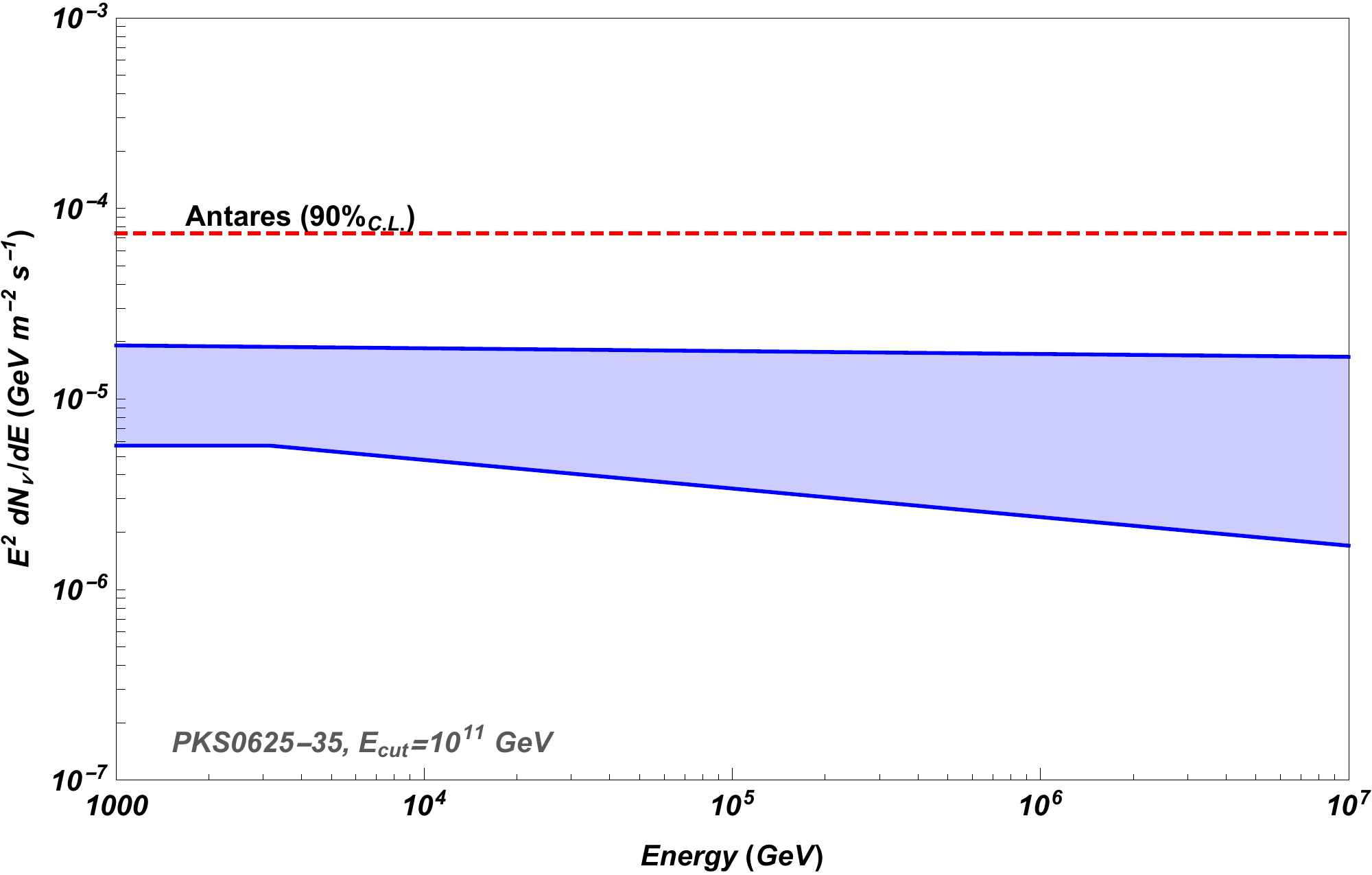} \\
\vspace{0.5cm}
\includegraphics[scale=0.65]{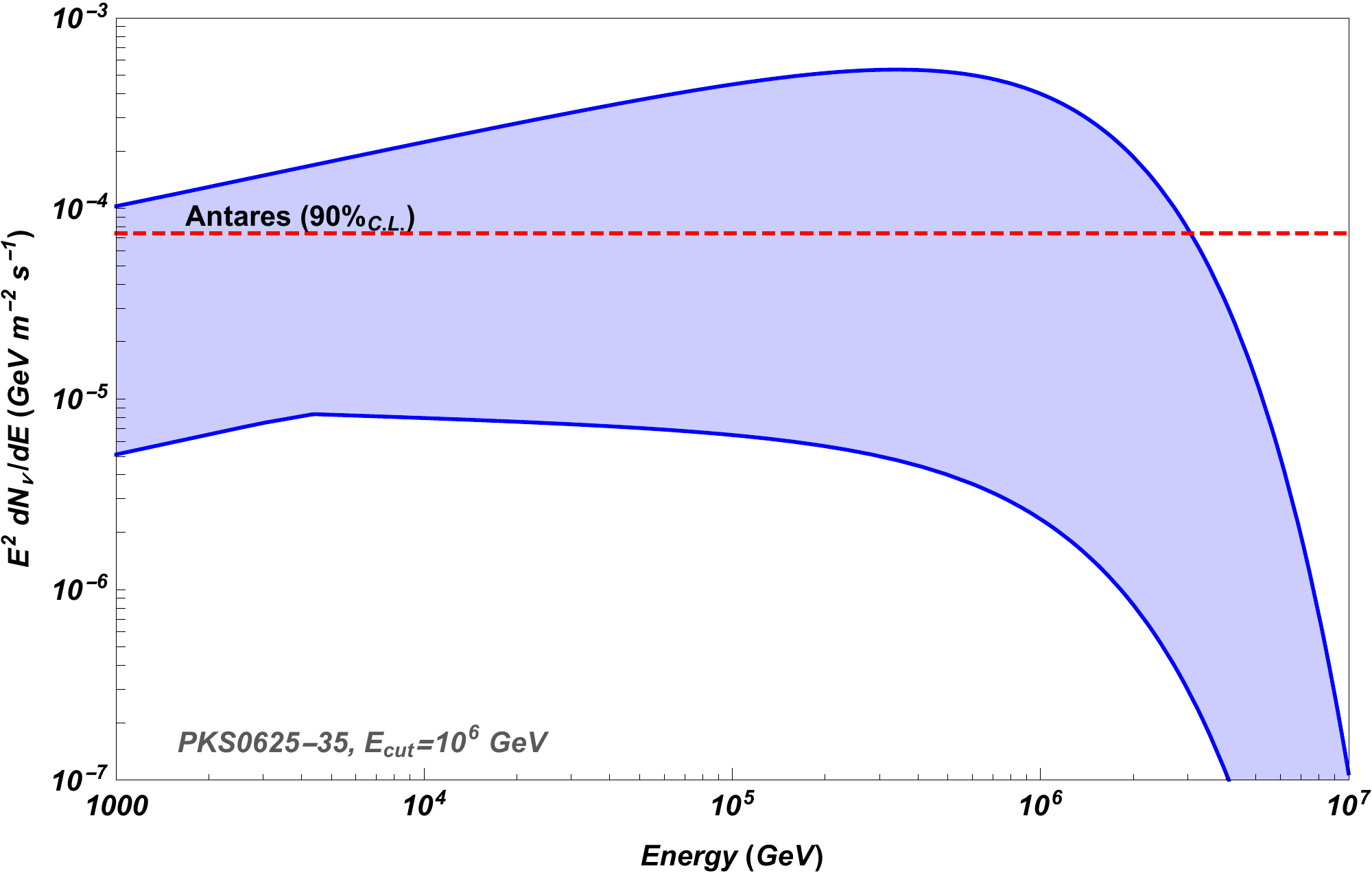}
\caption{As in Fig.~\ref{neutrinospecCenA}, but for the radio galaxy PKS 0625-35. Also shown is the (90\% confidence level) upper limit from this source as reported by the ANTARES Collaboration~\cite{Albert:2017ohr}.}
\label{neutrinospecPKS}
\end{figure}

\begin{figure}[h]
\includegraphics[scale=0.65]{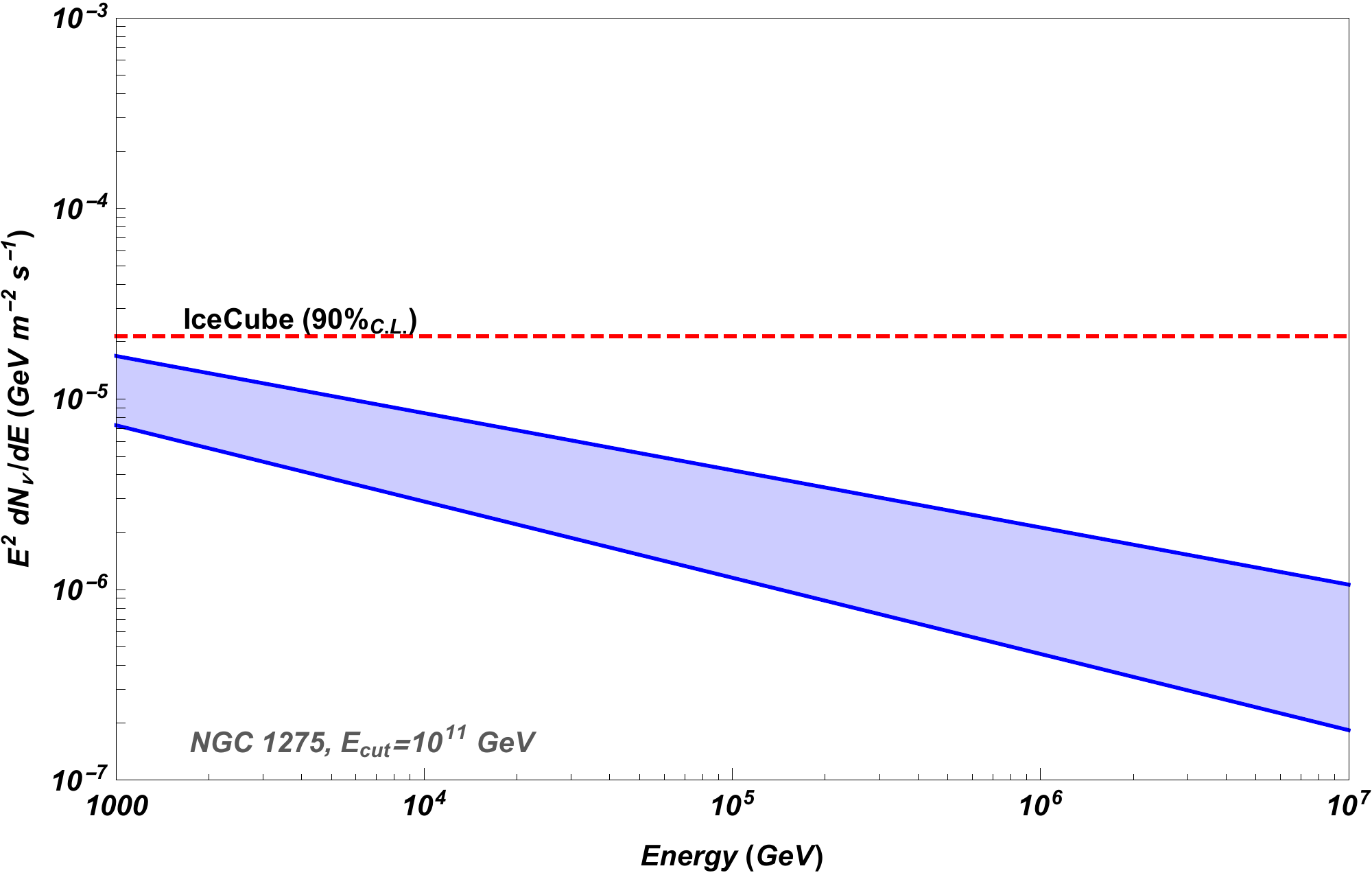} \\
\vspace{0.5cm}
\includegraphics[scale=0.65]{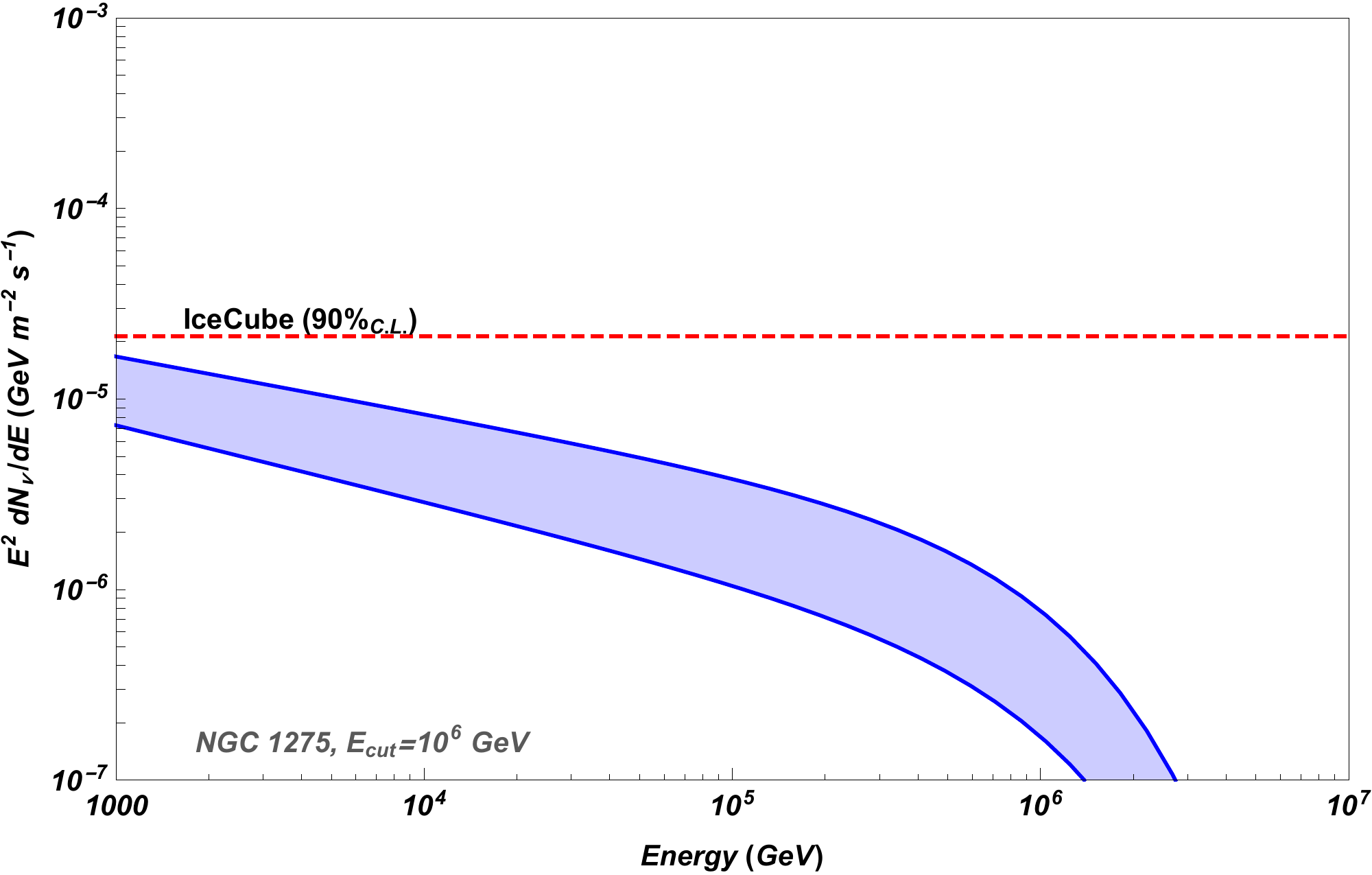}
\caption{As in Figs.~\ref{neutrinospecCenA}-\ref{neutrinospecPKS}, but for the radio galaxy NGC 1275. We also show the (90\% confidence level) upper limit from this source as reported by the IceCube Collaboration~\cite{Aartsen:2016oji}.}
\label{neutrinospecNGC}
\end{figure}

\begin{figure}[h]
\includegraphics[scale=0.65]{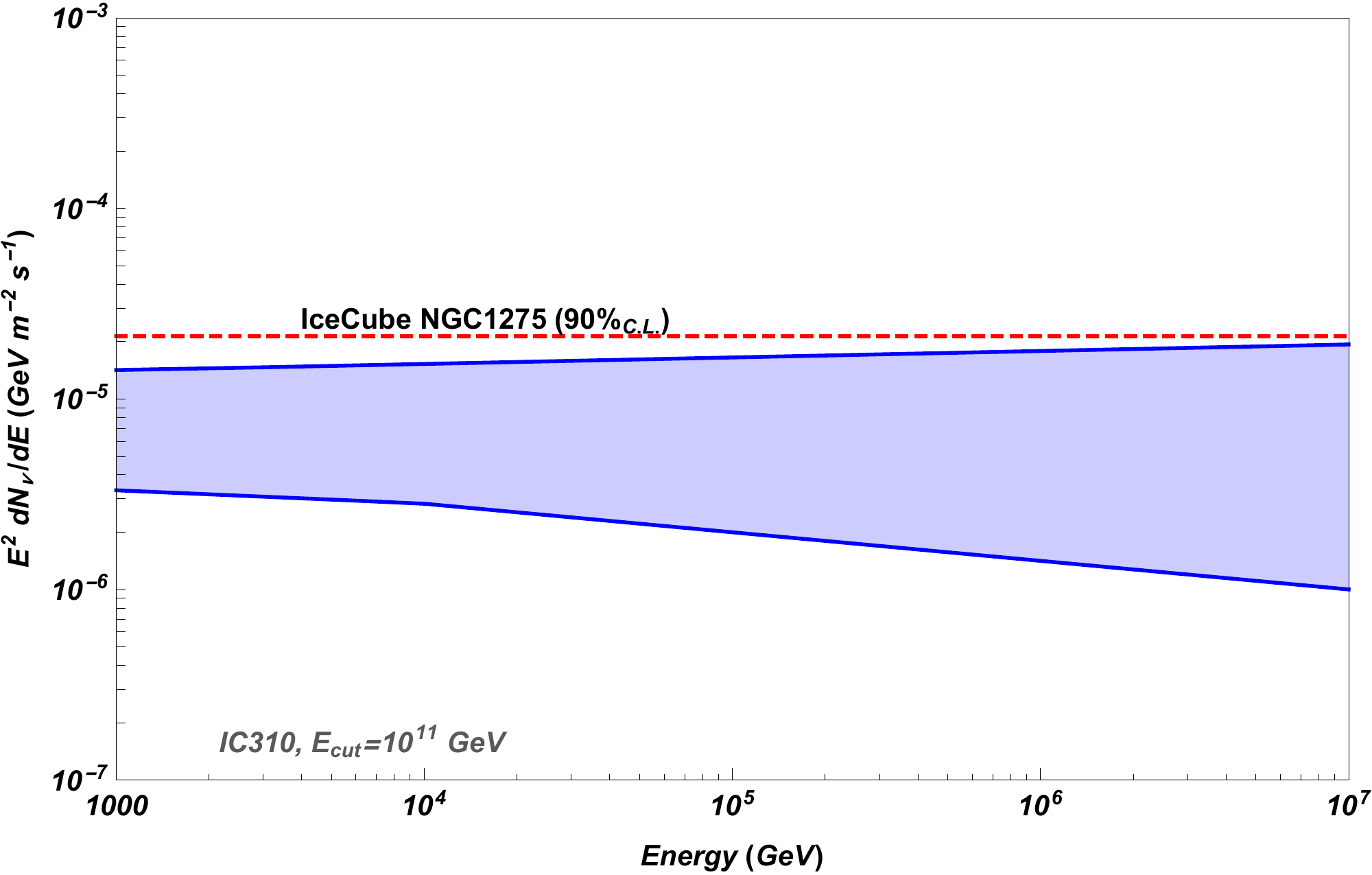} \\
\vspace{0.5cm}
\includegraphics[scale=0.65]{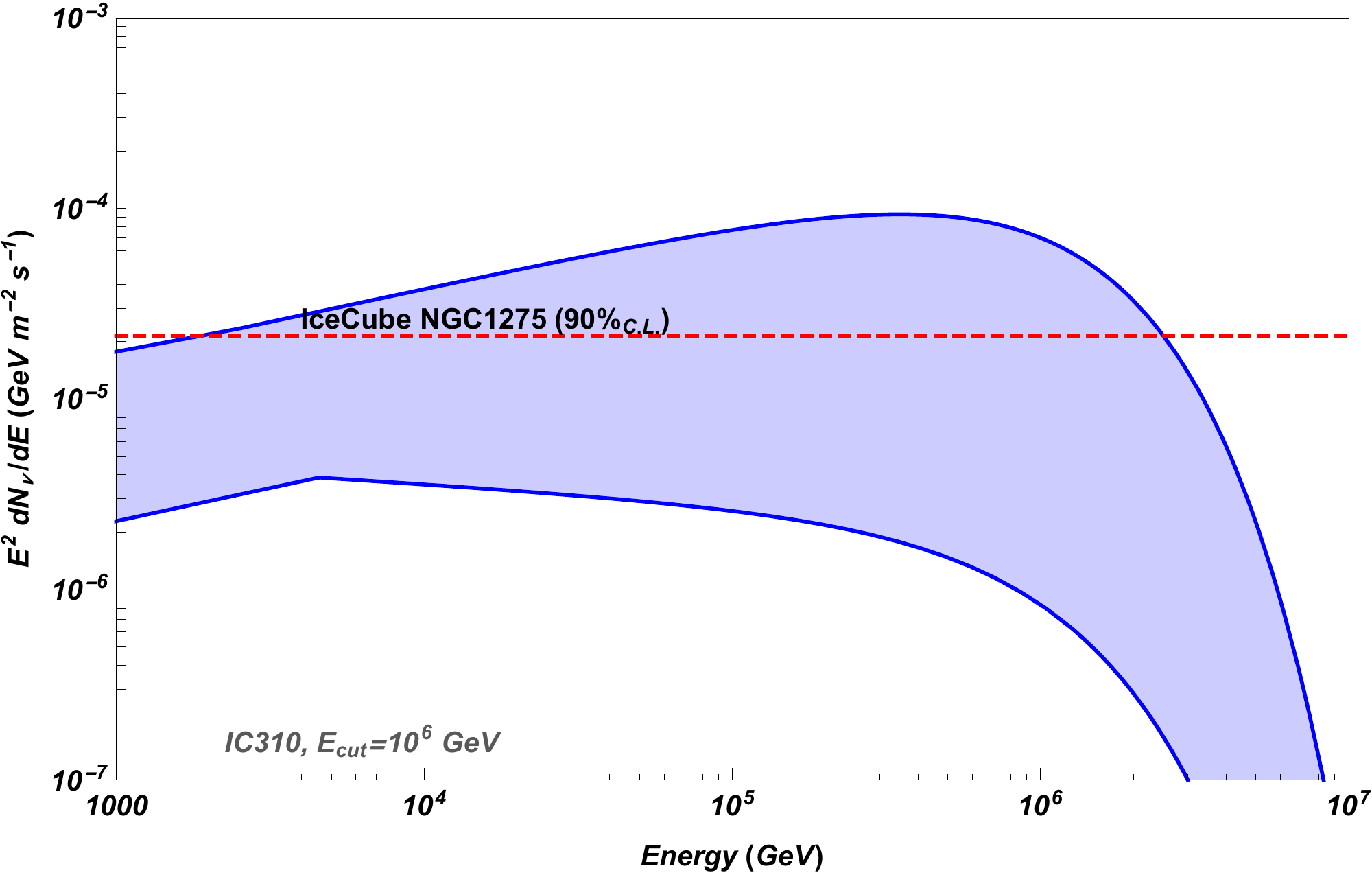}
\caption{As in Figs.~\ref{neutrinospecCenA}-\ref{neutrinospecNGC}, but for the radio galaxy IC 310. We also show the (90\% confidence level) upper limit from NGC 1275 as reported by the IceCube Collaboration~\cite{Aartsen:2016oji,Albert:2017ohr}, which is located $0.6^{\circ}$ away from IC 310 on the sky.}
\label{neutrinospecIC}
\end{figure}

From this relationship, we can utilize the injected spectra of gamma rays found to be favored by the fits described in Sec.~\ref{gammaresults} to predict the spectrum of high-energy neutrinos from the same collection of radio galaxies (Centaurus A, PKS 0625-35, NGC 1275 and IC 310). These results are shown in Figs.~\ref{neutrinospecCenA}-\ref{neutrinospecIC}. The band in each of these figures represents the predicted neutrino spectrum, after marginalizing over the normalization and spectral index of the injected gamma-ray spectrum (at the 1$\sigma$ level). In each figure, we show results assuming a maximum injected gamma-ray energy of $E_{\rm cut}=10^{11}$ GeV (upper frames) and $E_{\rm cut}=10^{6}$ GeV (lower frames). In Figs.~\ref{neutrinospecCenA}-\ref{neutrinospecNGC} we plot the current constraints on the flux from these sources, as presented by the IceCube and ANTARES Collaborations~\cite{Aartsen:2016oji,Albert:2017ohr,Aartsen:2017eiu,aartsen2014searches,adrian2016first}. In the case of IC 310 (Fig.~\ref{neutrinospecIC}) we show the upper limit from the direction of NGC 1275~\cite{Aartsen:2016oji}, which is located an angular distance of only $0.6^{\circ}$ from IC 310 on the sky (IceCube has not published a limit from the direction of IC 310 itself). We note that the locations of IceCube and ANTARES leads to a natural complementarity in which these detectors are most sensitive to point sources located in the northern and southern skies, respectively.

From the results presented in these figures, we conclude that the scenario considered here (namely, that in which the gamma-ray emission from these radio galaxies is generated largely through proton-proton collisions) is compatible with the current constraints from IceCube and ANTARES~\cite{Aartsen:2016oji,Albert:2017ohr}. That being said, the existing constraints are only slightly less stringent than that expected to be required to test this scenario. In particular, the current constraints on PKS 0625-35 and IC 310 each exclude a portion of the parameter space that would have otherwise been favored by the observed gamma-ray spectra. More generally, we predict neutrino fluxes from these sources that are within an order of magnitude or less of the current constraints.\footnote{The neutrino point source limits considered here have been derived assuming a spectrum of the form $dN_{\nu}/dE_{\nu} \propto  E_{\nu}^{-2}$.}

In light of these results, it appears that the scenario considered in this paper is within the reach of future, if not current, high-energy neutrino telescopes. Very roughly speaking, we expect that a stacked analysis of the nearest radio galaxies, utilizing the full existing dataset from IceCube and ANTARES, to be able to improve on the current (single source) sensitivity by a factor of a few. Furthermore, plans are currently under consideration for significant extensions of these experiments~\cite{Blaufuss:2015muc,Adrian-Martinez:2016fdl,Halzen:2003fi}, with the potential to definitively test this class of scenarios.  As the results of this study indicate, an order of magnitude improvement over the current sensitivity to TeV-PeV neutrinos should be sufficient to confirm or refute the well-motivated hypothesis that radio galaxies are responsible for IceCube's diffuse neutrino flux (along with the bulk of Fermi's IGRB).

\section{Summary and Conclusions}

Previous studies have demonstrated that the GeV-TeV isotropic gamma-ray background is dominate by emission from unresolved radio galaxies (\ie non-blazar active galaxies)~\cite{Hooper:2016gjy,DiMauro:2013xta}. Furthermore, if this gamma-ray emission is taken to be produced through the interactions of high-energy protons with gas, it has been shown that the corresponding neutrino flux is in good agreement with that reported by the IceCube Collaboration~\cite{Hooper:2016jls}. In light of these considerations, and given the evidence against gamma-ray bursts~\cite{Abbasi:2012zw,Aartsen:2016qcr}, blazars~\cite{Glusenkamp:2015jca,Ahlers:2014ioa}, and star-forming galaxies~\cite{Bechtol:2015uqb} as the primary sources of the observed neutrinos, radio galaxies now appear to be the most likely class of sources for IceCube's reported signal.

In this paper, we have considered the gamma-ray spectra from four nearby radio galaxies (Centaurus A, PKS 0625-35, NGC 1275 and IC 310), as measured in the $\sim$1-100 GeV range by Fermi, and in the $\sim$0.1-10 TeV range by HESS, VERITAS and/or MAGIC. From this information, we have determined the range of the injected gamma-ray intensity and spectral index that can account for these observations, after accounting for the effects of attenuation and contributions from electromagnetic cascades. We then calculated the predicted spectrum of high-energy neutrinos from these sources, assuming that their observed gamma-ray emission is generated primarily through cosmic-ray proton interactions with gas. 

Our predictions for the neutrino spectra from these four sources generally fall below the constraints currently placed by the IceCube and ANTARES Collaborations~\cite{Aartsen:2016oji,Albert:2017ohr}. That being said, the current constraints are typically only a factor of $\sim$1-10 above the fluxes predicted in this study. This suggests that a stacked analysis of radio galaxies making use of the complete data set from these neutrino telescopes could be sensitive to a significant fraction of the favored parameter space. Furthermore, future high-energy neutrinos telescopes with somewhat greater sensitivity to TeV-PeV neutrinos~\cite{Blaufuss:2015muc,Adrian-Martinez:2016fdl,Halzen:2003fi} could definitively test this scenario in the relatively near future.

In addition to neutrino observations, future gamma-ray observations will enable us to refine and test the predictions of this scenario. In particular, the upcoming Cherenkov Telescope Array (CTA) is expected to measure the very high-energy gamma-ray spectra from a number of radio galaxies with unprecedented precision. Depending on the strength of the intergalactic magnetic field, it is also possible that CTA could detect an extended component of gamma-ray emission from these sources, resulting from the deflection of electrons and positrons within electromagnetic cascades.

The detection of high-energy neutrinos from individual radio galaxies would provide a ``smoking gun'' for the acceleration of high-energy cosmic rays, likely in connection with the origin of the ultra high-energy cosmic ray spectrum. A unified picture appears to be emerging in which active galaxies accelerate cosmic ray protons and nuclei with a power-law index of $\sim$\,2. Through pion production, these cosmic rays then transfer on the order of 10\% of their energy into the production of pions, which decay to generate both the diffuse flux of neutrinos observed by IceCube, as well as a large fraction of Fermi's isotropic gamma-ray background.

\bigskip

\textbf{Acknowledgments.} We would like to thank Kohta Murase for helpful discussions. This manuscript has been authored by Fermi Research Alliance, LLC under Contract No. DE-AC02-07CH11359 with the U.S. Department of Energy, Office of Science, Office of High Energy Physics. The United States Government retains and the publisher, by accepting the article for publication, acknowledges that the United States Government retains a non-exclusive, paid-up, irrevocable, world-wide license to publish or reproduce the published form of this manuscript, or allow others to do so, for United States Government purposes. C. B. is supported by the National Science Foundation Graduate Research Fellowship under Grants No. DGE-1144082 and DGE-1746045.

\bibliography{RadioGalaxyRefs}

\providecommand{\href}[2]{#2}\begingroup\raggedright\begin{thebibliography}{100}

\bibitem{Aartsen:2013bka}
{\scshape IceCube} collaboration, M.~G. Aartsen et~al., \emph{{First
  observation of PeV-energy neutrinos with IceCube}},
  \href{http://dx.doi.org/10.1103/PhysRevLett.111.021103}{\emph{Phys. Rev.
  Lett.} {\bf 111} (2013) 021103}, [\href{http://arxiv.org/abs/1304.5356}{{\tt
  1304.5356}}].

\bibitem{Waxman:1997ti}
E.~Waxman and J.~N. Bahcall, \emph{{High-energy neutrinos from cosmological
  gamma-ray burst fireballs}},
  \href{http://dx.doi.org/10.1103/PhysRevLett.78.2292}{\emph{Phys. Rev. Lett.}
  {\bf 78} (1997) 2292--2295},
  [\href{http://arxiv.org/abs/astro-ph/9701231}{{\tt astro-ph/9701231}}].

\bibitem{Rachen:1998ir}
J.~P. Rachen and P.~Meszaros, \emph{{Cosmic rays and neutrinos from gamma-ray
  bursts}}, \href{http://dx.doi.org/10.1063/1.55402}{\emph{AIP Conf. Proc.}
  {\bf 428} (1997) 776}, [\href{http://arxiv.org/abs/astro-ph/9811266}{{\tt
  astro-ph/9811266}}].

\bibitem{Guetta:2003wi}
D.~Guetta, D.~Hooper, J.~Alvarez-Muniz, F.~Halzen and E.~Reuveni,
  \emph{{Neutrinos from individual gamma-ray bursts in the BATSE catalog}},
  \href{http://dx.doi.org/10.1016/S0927-6505(03)00211-1}{\emph{Astropart.
  Phys.} {\bf 20} (2004) 429--455},
  [\href{http://arxiv.org/abs/astro-ph/0302524}{{\tt astro-ph/0302524}}].

\bibitem{Stecker:1991vm}
F.~W. Stecker, C.~Done, M.~H. Salamon and P.~Sommers, \emph{{High-energy
  neutrinos from active galactic nuclei}},
  \href{http://dx.doi.org/10.1103/PhysRevLett.66.2697}{\emph{Phys. Rev. Lett.}
  {\bf 66} (1991) 2697--2700}.

\bibitem{Halzen:1997hw}
F.~Halzen and E.~Zas, \emph{{Neutrino fluxes from active galaxies: A Model
  independent estimate}},
  \href{http://dx.doi.org/10.1086/304741}{\emph{Astrophys. J.} {\bf 488} (1997)
  669--674}, [\href{http://arxiv.org/abs/astro-ph/9702193}{{\tt
  astro-ph/9702193}}].

\bibitem{Atoyan:2001ey}
A.~Atoyan and C.~D. Dermer, \emph{{High-energy neutrinos from photomeson
  processes in blazars}},
  \href{http://dx.doi.org/10.1103/PhysRevLett.87.221102}{\emph{Phys. Rev.
  Lett.} {\bf 87} (2001) 221102},
  [\href{http://arxiv.org/abs/astro-ph/0108053}{{\tt astro-ph/0108053}}].

\bibitem{Murase:2015ndr}
K.~Murase, \emph{{Active Galactic Nuclei as High-Energy Neutrino Sources}},
  \href{http://arxiv.org/abs/1511.01590}{{\tt 1511.01590}}.

\bibitem{Loeb:2006tw}
A.~Loeb and E.~Waxman, \emph{{The Cumulative background of high energy
  neutrinos from starburst galaxies}},
  \href{http://dx.doi.org/10.1088/1475-7516/2006/05/003}{\emph{JCAP} {\bf 0605}
  (2006) 003}, [\href{http://arxiv.org/abs/astro-ph/0601695}{{\tt
  astro-ph/0601695}}].

\bibitem{Becker:2007sv}
J.~K. Becker, \emph{{High-energy neutrinos in the context of multimessenger
  physics}}, \href{http://dx.doi.org/10.1016/j.physrep.2007.10.006}{\emph{Phys.
  Rept.} {\bf 458} (2008) 173--246},
  [\href{http://arxiv.org/abs/0710.1557}{{\tt 0710.1557}}].

\bibitem{Halzen:2002pg}
F.~Halzen and D.~Hooper, \emph{{High-energy neutrino astronomy: The Cosmic ray
  connection}}, \href{http://dx.doi.org/10.1088/0034-4885/65/7/201}{\emph{Rept.
  Prog. Phys.} {\bf 65} (2002) 1025--1078},
  [\href{http://arxiv.org/abs/astro-ph/0204527}{{\tt astro-ph/0204527}}].

\bibitem{Abbasi:2012zw}
{\scshape IceCube} collaboration, R.~Abbasi et~al., \emph{{An absence of
  neutrinos associated with cosmic-ray acceleration in $\gamma$-ray bursts}},
  \href{http://dx.doi.org/10.1038/nature11068}{\emph{Nature} {\bf 484} (2012)
  351--353}, [\href{http://arxiv.org/abs/1204.4219}{{\tt 1204.4219}}].

\bibitem{Aartsen:2016qcr}
{\scshape IceCube} collaboration, M.~G. Aartsen et~al., \emph{{An All-Sky
  Search for Three Flavors of Neutrinos from Gamma-Ray Bursts with the IceCube
  Neutrino Observatory}},
  \href{http://dx.doi.org/10.3847/0004-637X/824/2/115}{\emph{Astrophys. J.}
  {\bf 824} (2016) 115}, [\href{http://arxiv.org/abs/1601.06484}{{\tt
  1601.06484}}].

\bibitem{Murase:2013ffa}
K.~Murase and K.~Ioka, \emph{{TeV–PeV Neutrinos from Low-Power Gamma-Ray
  Burst Jets inside Stars}},
  \href{http://dx.doi.org/10.1103/PhysRevLett.111.121102}{\emph{Phys. Rev.
  Lett.} {\bf 111} (2013) 121102}, [\href{http://arxiv.org/abs/1306.2274}{{\tt
  1306.2274}}].

\bibitem{Tamborra:2015qza}
I.~Tamborra and S.~Ando, \emph{{Diffuse emission of high-energy neutrinos from
  gamma-ray burst fireballs}},
  \href{http://dx.doi.org/10.1088/1475-7516/2015/9/036,
  10.1088/1475-7516/2015/09/036}{\emph{JCAP} {\bf 1509} (2015) 036},
  [\href{http://arxiv.org/abs/1504.00107}{{\tt 1504.00107}}].

\bibitem{Senno:2015tsn}
N.~Senno, K.~Murase and P.~Meszaros, \emph{{Choked Jets and Low-Luminosity
  Gamma-Ray Bursts as Hidden Neutrino Sources}},
  \href{http://dx.doi.org/10.1103/PhysRevD.93.083003}{\emph{Phys. Rev.} {\bf
  D93} (2016) 083003}, [\href{http://arxiv.org/abs/1512.08513}{{\tt
  1512.08513}}].

\bibitem{Tamborra:2015fzv}
I.~Tamborra and S.~Ando, \emph{{Inspecting the supernova–gamma-ray-burst
  connection with high-energy neutrinos}},
  \href{http://dx.doi.org/10.1103/PhysRevD.93.053010}{\emph{Phys. Rev.} {\bf
  D93} (2016) 053010}, [\href{http://arxiv.org/abs/1512.01559}{{\tt
  1512.01559}}].

\bibitem{Glusenkamp:2015jca}
{\scshape IceCube} collaboration, T.~Glüsenkamp, \emph{{Analysis of the
  cumulative neutrino flux from Fermi-LAT blazar populations using 3 years of
  IceCube data}},  in \emph{{5th Roma International Conference on
  Astro-Particle physics (RICAP 14) Noto, Sicily, Italy, September 30-October
  3, 2014}}, 2015.
\newblock \href{http://arxiv.org/abs/1502.03104}{{\tt 1502.03104}}.

\bibitem{Ahlers:2014ioa}
M.~Ahlers and F.~Halzen, \emph{{Pinpointing Extragalactic Neutrino Sources in
  Light of Recent IceCube Observations}},
  \href{http://dx.doi.org/10.1103/PhysRevD.90.043005}{\emph{Phys. Rev.} {\bf
  D90} (2014) 043005}, [\href{http://arxiv.org/abs/1406.2160}{{\tt
  1406.2160}}].

\bibitem{TheFermi-LAT:2015ykq}
{\scshape Fermi-LAT} collaboration, M.~Ackermann et~al., \emph{{Resolving the
  Extragalactic $\gamma$-Ray Background above 50 GeV with the Fermi Large Area
  Telescope}},
  \href{http://dx.doi.org/10.1103/PhysRevLett.116.151105}{\emph{Phys. Rev.
  Lett.} {\bf 116} (2016) 151105}, [\href{http://arxiv.org/abs/1511.00693}{{\tt
  1511.00693}}].

\bibitem{Bechtol:2015uqb}
K.~Bechtol, M.~Ahlers, M.~Di~Mauro, M.~Ajello and J.~Vandenbroucke,
  \emph{{Evidence against star-forming galaxies as the dominant source of
  IceCube neutrinos}},  \href{http://arxiv.org/abs/1511.00688}{{\tt
  1511.00688}}.

\bibitem{Murase:2015xka}
K.~Murase, D.~Guetta and M.~Ahlers, \emph{{Hidden Cosmic-Ray Accelerators as an
  Origin of TeV-PeV Cosmic Neutrinos}},
  \href{http://dx.doi.org/10.1103/PhysRevLett.116.071101}{\emph{Phys. Rev.
  Lett.} {\bf 116} (2016) 071101}, [\href{http://arxiv.org/abs/1509.00805}{{\tt
  1509.00805}}].

\bibitem{Chakraborty:2016mvc}
S.~Chakraborty and I.~Izaguirre, \emph{{Star-forming galaxies as the origin of
  IceCube neutrinos: Reconciliation with Fermi-LAT gamma rays}},
  \href{http://arxiv.org/abs/1607.03361}{{\tt 1607.03361}}.

\bibitem{Xiao:2016rvd}
D.~Xiao, P.~Mészáros, K.~Murase and Z.-g. Dai, \emph{{Revisiting the
  Contributions of Supernova and Hypernova Remnants to the Diffuse High-energy
  Backgrounds: Constraints on Very High Redshift Injection}},
  \href{http://dx.doi.org/10.3847/0004-637X/826/2/133}{\emph{Astrophys. J.}
  {\bf 826} (2016) 133}, [\href{http://arxiv.org/abs/1604.08131}{{\tt
  1604.08131}}].

\bibitem{Murase:2013rfa}
K.~Murase, M.~Ahlers and B.~C. Lacki, \emph{{Testing the Hadronuclear Origin of
  PeV Neutrinos Observed with IceCube}},
  \href{http://dx.doi.org/10.1103/PhysRevD.88.121301}{\emph{Phys. Rev.} {\bf
  D88} (2013) 121301}, [\href{http://arxiv.org/abs/1306.3417}{{\tt
  1306.3417}}].

\bibitem{Hooper:2016jls}
D.~Hooper, \emph{{A Case for Radio Galaxies as the Sources of IceCube's
  Astrophysical Neutrino Flux}},
  \href{http://dx.doi.org/10.1088/1475-7516/2016/09/002}{\emph{JCAP} {\bf 1609}
  (2016) 002}, [\href{http://arxiv.org/abs/1605.06504}{{\tt 1605.06504}}].

\bibitem{Chang:2016ljk}
X.-C. Chang, R.-Y. Liu and X.-Y. Wang, \emph{{How far are the sources of
  IceCube neutrinos? Constraints from the diffuse TeV gamma-ray background}},
  \href{http://arxiv.org/abs/1602.06625}{{\tt 1602.06625}}.

\bibitem{Kalashev:2014vra}
O.~E. Kalashev and S.~V. Troitsky, \emph{{IceCube astrophysical neutrinos
  without a spectral cutoff and $10^{15}$–$10^{17}$ eV cosmic gamma
  radiation}}, \href{http://dx.doi.org/10.1134/S0021364014240072}{\emph{JETP
  Lett.} {\bf 100} (2015) 761--765},
  [\href{http://arxiv.org/abs/1410.2600}{{\tt 1410.2600}}].

\bibitem{Giacinti:2015pya}
G.~Giacinti, M.~Kachelrieß, O.~Kalashev, A.~Neronov and D.~Semikoz,
  \emph{{Unified model for cosmic rays above 10$^{17}$  eV and the diffuse
  gamma-ray and neutrino backgrounds}},
  \href{http://dx.doi.org/10.1103/PhysRevD.92.083016}{\emph{Phys. Rev.} {\bf
  D92} (2015) 083016}, [\href{http://arxiv.org/abs/1507.07534}{{\tt
  1507.07534}}].

\bibitem{Tjus:2014dna}
J.~Becker~Tjus, B.~Eichmann, F.~Halzen, A.~Kheirandish and S.~Saba,
  \emph{{High-energy neutrinos from radio galaxies}},
  \href{http://dx.doi.org/10.1103/PhysRevD.89.123005}{\emph{Phys. Rev.} {\bf
  D89} (2014) 123005}, [\href{http://arxiv.org/abs/1406.0506}{{\tt
  1406.0506}}].

\bibitem{Murase:2014foa}
K.~Murase, Y.~Inoue and C.~D. Dermer, \emph{{Diffuse Neutrino Intensity from
  the Inner Jets of Active Galactic Nuclei: Impacts of External Photon Fields
  and the Blazar Sequence}},
  \href{http://dx.doi.org/10.1103/PhysRevD.90.023007}{\emph{Phys. Rev.} {\bf
  D90} (2014) 023007}, [\href{http://arxiv.org/abs/1403.4089}{{\tt
  1403.4089}}].

\bibitem{Kimura:2014jba}
S.~S. Kimura, K.~Murase and K.~Toma, \emph{{Neutrino and Cosmic-Ray Emission
  and Cumulative Background from Radiatively Inefficient Accretion Flows in
  Low-Luminosity Active Galactic Nuclei}},
  \href{http://dx.doi.org/10.1088/0004-637X/806/2/159}{\emph{Astrophys. J.}
  {\bf 806} (2015) 159}, [\href{http://arxiv.org/abs/1411.3588}{{\tt
  1411.3588}}].

\bibitem{Murase:2016gly}
K.~Murase and E.~Waxman, \emph{{Constraining High-Energy Cosmic Neutrino
  Sources: Implications and Prospects}},
  \href{http://dx.doi.org/10.1103/PhysRevD.94.103006}{\emph{Phys. Rev.} {\bf
  D94} (2016) 103006}, [\href{http://arxiv.org/abs/1607.01601}{{\tt
  1607.01601}}].

\bibitem{Sahu:2012wv}
S.~Sahu, B.~Zhang and N.~Fraija, \emph{{Hadronic-Origin TeV gamma-Rays and
  Ultra-High Energy Cosmic Rays from Centaurus A}},
  \href{http://dx.doi.org/10.1103/PhysRevD.85.043012}{\emph{Phys. Rev.} {\bf
  D85} (2012) 043012}, [\href{http://arxiv.org/abs/1201.4191}{{\tt
  1201.4191}}].

\bibitem{Marinelli:2014mva}
A.~Marinelli, N.~Fraija and B.~Patricelli, \emph{{The Hadronic Picture of the
  Radiogalaxy M87}},  \href{http://arxiv.org/abs/1410.8549}{{\tt 1410.8549}}.

\bibitem{Fraija:2017jok}
N.~Fraija, E.~Aguilar-Ruiz, A.~Galván-Gámez, A.~Marinelli and J.~A. de~Diego,
  \emph{{Study of the PeV Neutrino, $\gamma$-rays and UHECRs around The Lobes
  of Centaurus A}},  \href{http://arxiv.org/abs/1709.05766}{{\tt 1709.05766}}.

\bibitem{Fraija:2012na}
N.~Fraija, M.~M. Gonzalez and M.~Perez, \emph{{Hadronic processes as origin of
  TeV emission in Fanaroff-Riley Class I: Cen A, M87 and NGC1275}}, {\emph{PoS}
  {\bf GRB2012} (2012) 131}, [\href{http://arxiv.org/abs/1212.4424}{{\tt
  1212.4424}}].

\bibitem{Ackermann:2014usa}
{\scshape Fermi-LAT} collaboration, M.~Ackermann et~al., \emph{{The spectrum of
  isotropic diffuse gamma-ray emission between 100 MeV and 820 GeV}},
  \href{http://dx.doi.org/10.1088/0004-637X/799/1/86}{\emph{Astrophys. J.} {\bf
  799} (2015) 86}, [\href{http://arxiv.org/abs/1410.3696}{{\tt 1410.3696}}].

\bibitem{Hooper:2016gjy}
D.~Hooper, T.~Linden and A.~Lopez, \emph{{Radio Galaxies Dominate the
  High-Energy Diffuse Gamma-Ray Background}},
  \href{http://arxiv.org/abs/1604.08505}{{\tt 1604.08505}}.

\bibitem{Linden:2016fdd}
T.~Linden, \emph{{Star-Forming Galaxies Significantly Contribute to the
  Isotropic Gamma-Ray Background}},
  \href{http://arxiv.org/abs/1612.03175}{{\tt 1612.03175}}.

\bibitem{Tamborra:2014xia}
I.~Tamborra, S.~Ando and K.~Murase, \emph{{Star-forming galaxies as the origin
  of diffuse high-energy backgrounds: Gamma-ray and neutrino connections, and
  implications for starburst history}},
  \href{http://dx.doi.org/10.1088/1475-7516/2014/09/043}{\emph{JCAP} {\bf 1409}
  (2014) 043}, [\href{http://arxiv.org/abs/1404.1189}{{\tt 1404.1189}}].

\bibitem{Ackermann:2012vca}
{\scshape Fermi-LAT} collaboration, M.~Ackermann et~al., \emph{{GeV
  Observations of Star-forming Galaxies with \textit{Fermi} LAT}},
  \href{http://dx.doi.org/10.1088/0004-637X/755/2/164}{\emph{Astrophys. J.}
  {\bf 755} (2012) 164}, [\href{http://arxiv.org/abs/1206.1346}{{\tt
  1206.1346}}].

\bibitem{Stecker:2010di}
F.~W. Stecker and T.~M. Venters, \emph{{Components of the Extragalactic Gamma
  Ray Background}},
  \href{http://dx.doi.org/10.1088/0004-637X/736/1/40}{\emph{Astrophys. J.} {\bf
  736} (2011) 40}, [\href{http://arxiv.org/abs/1012.3678}{{\tt 1012.3678}}].

\bibitem{DiMauro:2013xta}
M.~Di~Mauro, F.~Calore, F.~Donato, M.~Ajello and L.~Latronico, \emph{{Diffuse
  $\gamma$-ray emission from misaligned active galactic nuclei}},
  \href{http://dx.doi.org/10.1088/0004-637X/780/2/161}{\emph{Astrophys. J.}
  {\bf 780} (2014) 161}, [\href{http://arxiv.org/abs/1304.0908}{{\tt
  1304.0908}}].

\bibitem{Cuoco:2012yf}
A.~Cuoco, E.~Komatsu and J.~M. Siegal-Gaskins, \emph{{Joint anisotropy and
  source count constraints on the contribution of blazars to the diffuse
  gamma-ray background}},
  \href{http://dx.doi.org/10.1103/PhysRevD.86.063004}{\emph{Phys. Rev.} {\bf
  D86} (2012) 063004}, [\href{http://arxiv.org/abs/1202.5309}{{\tt
  1202.5309}}].

\bibitem{Harding:2012gk}
J.~P. Harding and K.~N. Abazajian, \emph{{Models of the Contribution of Blazars
  to the Anisotropy of the Extragalactic Diffuse Gamma-ray Background}},
  \href{http://dx.doi.org/10.1088/1475-7516/2012/11/026}{\emph{JCAP} {\bf 1211}
  (2012) 026}, [\href{http://arxiv.org/abs/1206.4734}{{\tt 1206.4734}}].

\bibitem{Ajello:2011zi}
M.~Ajello et~al., \emph{{The Luminosity Function of Fermi-detected
  Flat-Spectrum Radio Quasars}},
  \href{http://dx.doi.org/10.1088/0004-637X/751/2/108}{\emph{Astrophys. J.}
  {\bf 751} (2012) 108}, [\href{http://arxiv.org/abs/1110.3787}{{\tt
  1110.3787}}].

\bibitem{Ajello:2013lka}
M.~Ajello et~al., \emph{{The Cosmic Evolution of Fermi BL Lacertae Objects}},
  \href{http://dx.doi.org/10.1088/0004-637X/780/1/73}{\emph{Astrophys. J.} {\bf
  780} (2014) 73}, [\href{http://arxiv.org/abs/1310.0006}{{\tt 1310.0006}}].

\bibitem{Zandanel:2014pva}
F.~Zandanel, I.~Tamborra, S.~Gabici and S.~Ando, \emph{{High-energy gamma-ray
  and neutrino backgrounds from clusters of galaxies and radio constraints}},
  \href{http://dx.doi.org/10.1051/0004-6361/201425249}{\emph{Astron.
  Astrophys.} {\bf 578} (2015) A32},
  [\href{http://arxiv.org/abs/1410.8697}{{\tt 1410.8697}}].

\bibitem{Taylor:2015rla}
A.~M. Taylor, M.~Ahlers and D.~Hooper, \emph{{Indications of Negative Evolution
  for the Sources of the Highest Energy Cosmic Rays}},
  \href{http://dx.doi.org/10.1103/PhysRevD.92.063011}{\emph{Phys. Rev.} {\bf
  D92} (2015) 063011}, [\href{http://arxiv.org/abs/1505.06090}{{\tt
  1505.06090}}].

\bibitem{Ahlers:2011sd}
M.~Ahlers and J.~Salvado, \emph{{Cosmogenic gamma-rays and the composition of
  cosmic rays}},
  \href{http://dx.doi.org/10.1103/PhysRevD.84.085019}{\emph{Phys. Rev.} {\bf
  D84} (2011) 085019}, [\href{http://arxiv.org/abs/1105.5113}{{\tt
  1105.5113}}].

\bibitem{Ackermann:2015tah}
{\scshape Fermi-LAT} collaboration, M.~Ackermann et~al., \emph{{Limits on Dark
  Matter Annihilation Signals from the Fermi LAT 4-year Measurement of the
  Isotropic Gamma-Ray Background}},
  \href{http://dx.doi.org/10.1088/1475-7516/2015/09/008}{\emph{JCAP} {\bf 1509}
  (2015) 008}, [\href{http://arxiv.org/abs/1501.05464}{{\tt 1501.05464}}].

\bibitem{DiMauro:2015tfa}
M.~Di~Mauro and F.~Donato, \emph{{Composition of the Fermi-LAT isotropic
  gamma-ray background intensity: Emission from extragalactic point sources and
  dark matter annihilations}},
  \href{http://dx.doi.org/10.1103/PhysRevD.91.123001}{\emph{Phys. Rev.} {\bf
  D91} (2015) 123001}, [\href{http://arxiv.org/abs/1501.05316}{{\tt
  1501.05316}}].

\bibitem{Ajello:2015mfa}
M.~Ajello et~al., \emph{{The Origin of the Extragalactic Gamma-Ray Background
  and Implications for Dark-Matter Annihilation}},
  \href{http://dx.doi.org/10.1088/2041-8205/800/2/L27}{\emph{Astrophys. J.}
  {\bf 800} (2015) L27}, [\href{http://arxiv.org/abs/1501.05301}{{\tt
  1501.05301}}].

\bibitem{Cholis:2013ena}
I.~Cholis, D.~Hooper and S.~D. McDermott, \emph{{Dissecting the Gamma-Ray
  Background in Search of Dark Matter}},
  \href{http://dx.doi.org/10.1088/1475-7516/2014/02/014}{\emph{JCAP} {\bf 1402}
  (2014) 014}, [\href{http://arxiv.org/abs/1312.0608}{{\tt 1312.0608}}].

\bibitem{Fornasa:2015qua}
M.~Fornasa and M.~A. Sánchez-Conde, \emph{{The nature of the Diffuse Gamma-Ray
  Background}},
  \href{http://dx.doi.org/10.1016/j.physrep.2015.09.002}{\emph{Phys. Rept.}
  {\bf 598} (2015) 1--58}, [\href{http://arxiv.org/abs/1502.02866}{{\tt
  1502.02866}}].

\bibitem{DiMauro:2016cbj}
{\scshape Fermi-LAT} collaboration, M.~Di~Mauro, \emph{{The origin of the
  Fermi-LAT $\gamma$-ray background}},  in \emph{{14th Marcel Grossmann Meeting
  on Recent Developments in Theoretical and Experimental General Relativity,
  Astrophysics, and Relativistic Field Theories (MG14) Rome, Italy, July 12-18,
  2015}}, 2016.
\newblock \href{http://arxiv.org/abs/1601.04323}{{\tt 1601.04323}}.

\bibitem{Cavadini:2011ig}
M.~Cavadini, R.~Salvaterra and F.~Haardt, \emph{{A New model for the
  extragalactic $\gamma$-ray background}},
  \href{http://arxiv.org/abs/1105.4613}{{\tt 1105.4613}}.

\bibitem{Siegal-Gaskins:2013tga}
J.~M. Siegal-Gaskins, \emph{{Separating astrophysical sources from indirect
  dark matter signals}},  in \emph{{Sackler Colloquium: Dark Matter Universe:
  On the Threshhold of Discovery Irvine, USA, October 18-20, 2012}}, 2013.
\newblock \href{http://arxiv.org/abs/1308.2228}{{\tt 1308.2228}}.
\newblock \href{http://dx.doi.org/10.1073/pnas.1315181111,
  10.1073/pnas.1516944112}{DOI}.

\bibitem{Xia:2015wka}
J.-Q. Xia, A.~Cuoco, E.~Branchini and M.~Viel, \emph{{Tomography of the
  Fermi-lat $\gamma$-ray Diffuse Extragalactic Signal via Cross Correlations
  With Galaxy Catalogs}},
  \href{http://dx.doi.org/10.1088/0067-0049/217/1/15}{\emph{Astrophys. J.
  Suppl.} {\bf 217} (2015) 15}, [\href{http://arxiv.org/abs/1503.05918}{{\tt
  1503.05918}}].

\bibitem{Cuoco:2015rfa}
A.~Cuoco, J.-Q. Xia, M.~Regis, E.~Branchini, N.~Fornengo and M.~Viel,
  \emph{{Dark Matter Searches in the Gamma-ray Extragalactic Background via
  Cross-correlations With Galaxy Catalogs}},
  \href{http://dx.doi.org/10.1088/0067-0049/221/2/29}{\emph{Astrophys. J.
  Suppl.} {\bf 221} (2015) 29}, [\href{http://arxiv.org/abs/1506.01030}{{\tt
  1506.01030}}].

\bibitem{Shirasaki:2014noa}
M.~Shirasaki, S.~Horiuchi and N.~Yoshida, \emph{{Cross-Correlation of Cosmic
  Shear and Extragalactic Gamma-ray Background: Constraints on the Dark Matter
  Annihilation Cross-Section}},
  \href{http://dx.doi.org/10.1103/PhysRevD.90.063502}{\emph{Phys. Rev.} {\bf
  D90} (2014) 063502}, [\href{http://arxiv.org/abs/1404.5503}{{\tt
  1404.5503}}].

\bibitem{Shirasaki:2015nqp}
M.~Shirasaki, S.~Horiuchi and N.~Yoshida, \emph{{Cross-Correlation of the
  Extragalactic Gamma-ray Background with Luminous Red Galaxies}},
  \href{http://dx.doi.org/10.1103/PhysRevD.92.123540}{\emph{Phys. Rev.} {\bf
  D92} (2015) 123540}, [\href{http://arxiv.org/abs/1511.07092}{{\tt
  1511.07092}}].

\bibitem{Ando:2015bva}
S.~Ando, I.~Tamborra and F.~Zandanel, \emph{{Tomographic Constraints on
  High-Energy Neutrinos of Hadronuclear Origin}},
  \href{http://dx.doi.org/10.1103/PhysRevLett.115.221101}{\emph{Phys. Rev.
  Lett.} {\bf 115} (2015) 221101}, [\href{http://arxiv.org/abs/1509.02444}{{\tt
  1509.02444}}].

\bibitem{Aharonian:2009xn}
{\scshape H.E.S.S.} collaboration, F.~Aharonian et~al., \emph{{Discovery of
  very high energy gamma-ray emission from Centaurus A with H.E.S.S}},
  \href{http://dx.doi.org/10.1088/0004-637X/695/1/L40}{\emph{Astrophys. J.}
  {\bf 695} (2009) L40--L44}, [\href{http://arxiv.org/abs/0903.1582}{{\tt
  0903.1582}}].

\bibitem{Dyrda:2015hxa}
{\scshape H.E.S.S.} collaboration, M.~Dyrda, A.~Wierzcholska, O.~Hervet,
  R.~Moderski, M.~Janiak, M.~Ostrowski et~al., \emph{{Discovery of VHE
  gamma-rays from the radio galaxy PKS 0625-354 with H.E.S.S}}, {\emph{PoS}
  {\bf ICRC2015} (2016) 801}, [\href{http://arxiv.org/abs/1509.06851}{{\tt
  1509.06851}}].

\bibitem{Ahnen:2016qkt}
{\scshape MAGIC} collaboration, M.~L. Ahnen et~al., \emph{{Deep observation of
  the NGC 1275 region with MAGIC: search of diffuse $\gamma$-ray emission
  from cosmic rays in the Perseus cluster}},
  \href{http://dx.doi.org/10.1051/0004-6361/201527846}{\emph{Astron.
  Astrophys.} {\bf 589} (2016) A33},
  [\href{http://arxiv.org/abs/1602.03099}{{\tt 1602.03099}}].

\bibitem{2017ATel.9931....1M}
R.~{Mukherjee} and {VERITAS Collaboration}, \emph{{VERITAS detection of the
  radio galaxy NGC 1275 with elevated very-high-energy gamma-ray emission}},
  {\emph{The Astronomer's Telegram} {\bf 9931} (Jan., 2017) }.

\bibitem{2016ATel.9690....1M}
R.~{Mukherjee} and {VERITAS Collaboration}, \emph{{VERITAS detection of the
  radio galaxy NGC 1275 with elevated very-high-energy gamma-ray emission}},
  {\emph{The Astronomer's Telegram} {\bf 9690} (Oct., 2016) }.

\bibitem{Aleksic:2013bya}
{\scshape MAGIC} collaboration, J.~Aleksić et~al., \emph{{Rapid and multiband
  variability of the TeV bright active nucleus of the galaxy IC 310}},
  \href{http://dx.doi.org/10.1051/0004-6361/201321938}{\emph{Astron.
  Astrophys.} {\bf 563} (2014) A91},
  [\href{http://arxiv.org/abs/1305.5147}{{\tt 1305.5147}}].

\bibitem{TheFermi-LAT:2015hja}
{\scshape Fermi-LAT Collaboration} collaboration, F.~Acero et~al., \emph{{Fermi
  Large Area Telescope Third Source Catalog}},
  \href{http://arxiv.org/abs/1501.02003}{{\tt 1501.02003}}.

\bibitem{Acciari:2008ah}
V.~A. Acciari et~al., \emph{{Observation of gamma-ray emission from the galaxy
  M87 above 250 GeV with VERITAS}},
  \href{http://dx.doi.org/10.1086/587458}{\emph{Astrophys. J.} {\bf 679} (2008)
  397}, [\href{http://arxiv.org/abs/0802.1951}{{\tt 0802.1951}}].

\bibitem{Acciari:2009rs}
{\scshape VERTIAS, HESS, MAGIC} collaboration, V.~A. Acciari et~al.,
  \emph{{Radio Imaging of the Very-High-Energy Gamma-Ray Emission Region in the
  Central Engine of a Radio Galaxy}},
  \href{http://dx.doi.org/10.1126/science.1175406}{\emph{Science} {\bf 325}
  (2009) 444--448}, [\href{http://arxiv.org/abs/0908.0511}{{\tt 0908.0511}}].

\bibitem{Galante:2009ie}
{\scshape VERITAS} collaboration, N.~Galante, \emph{{VERITAS Observations of
  Radio Galaxies}},  \href{http://arxiv.org/abs/0912.3850}{{\tt 0912.3850}}.

\bibitem{:2012uma}
{\scshape MAGIC} collaboration, J.~Aleksic et~al., \emph{{MAGIC observations of
  the giant radio galaxy M87 in a low-emission state between 2005 and 2007}},
  \href{http://dx.doi.org/10.1051/0004-6361/201117827}{\emph{Astron.
  Astrophys.} {\bf 544} (2012) A96},
  [\href{http://arxiv.org/abs/1207.2147}{{\tt 1207.2147}}].

\bibitem{Aleksic:2013kaa}
{\scshape MAGIC} collaboration, J.~Aleksić et~al., \emph{{Contemporaneous
  observations of the radio galaxy NGC 1275 from radio to very high energy
  $\gamma$-rays}},
  \href{http://dx.doi.org/10.1051/0004-6361/201322951}{\emph{Astron.
  Astrophys.} {\bf 564} (2014) A5}, [\href{http://arxiv.org/abs/1310.8500}{{\tt
  1310.8500}}].

\bibitem{Rieger2017}
F.~M. Rieger, F.~A. Aharonian, W.~Hofmann and F.~M. Rieger, \emph{Gamma-rays
  from non-blazar agn},  in \emph{AIP Conference Proceedings}, vol.~1792,
  p.~020008, AIP Publishing, 2017.

\bibitem{Aharonian2009}
F.~Aharonian, A.~Akhperjanian, G.~Anton, U.~B. De~Almeida, A.~Bazer-Bachi,
  Y.~Becherini et~al., \emph{Discovery of very high energy $\gamma$-ray
  emission from centaurus a with hess}, {\emph{The Astrophysical Journal
  Letters} {\bf 695} (2009) L40}.

\bibitem{yu2015high}
A.~P.-Y. Yu, J.~Lim, Y.~Ohyama, J.~C.-C. Chan and T.~Broadhurst, \emph{The
  high-velocity system: infall of a giant low-surface-brightness galaxy toward
  the center of the perseus cluster}, {\emph{The Astrophysical Journal} {\bf
  814} (2015) 101}.

\bibitem{gillmon2004}
K.~Gillmon, J.~Sanders and A.~Fabian, \emph{An x-ray absorption analysis of the
  high-velocity system in ngc 1275}, {\emph{Monthly Notices of the Royal
  Astronomical Society} {\bf 348} (2004) 159--164}.

\bibitem{brown2011high}
A.~M. Brown and J.~Adams, \emph{High-energy $\gamma$-ray properties of the
  fanaroff--riley type i radio galaxy ngc 1275}, {\emph{Monthly Notices of the
  Royal Astronomical Society} {\bf 413} (2011) 2785--2790}.

\bibitem{Conselice2001}
C.~J. Conselice, J.~S. Gallagher~III and R.~F. Wyse, \emph{On the nature of the
  ngc 1275 system}, {\emph{The Astronomical Journal} {\bf 122} (2001) 2281}.

\bibitem{aleksic2014contemporaneous}
J.~Aleksi{\'c}, S.~Ansoldi, L.~Antonelli, P.~Antoranz, A.~Babic, P.~Bangale
  et~al., \emph{Contemporaneous observations of the radio galaxy ngc 1275 from
  radio to very high energy $\gamma$-rays}, {\emph{Astronomy \& Astrophysics}
  {\bf 564} (2014) A5}.

\bibitem{aleksic2014}
J.~Aleksi{\'c}, S.~Ansoldi, L.~Antonelli, P.~Antoranz, A.~Babic, P.~Bangale
  et~al., \emph{Black hole lightning due to particle acceleration at subhorizon
  scales}, {\emph{Science} {\bf 346} (2014) 1080--1084}.

\bibitem{abramowski2012}
A.~Abramowski, F.~Acero, F.~Aharonian, A.~Akhperjanian, G.~Anton, A.~Balzer
  et~al., \emph{The 2010 very high energy $\gamma$-ray flare and 10 years of
  multi-wavelength observations of m 87}, {\emph{The Astrophysical Journal}
  {\bf 746} (2012) 151}.

\bibitem{albert2008}
J.~Albert, E.~Aliu, H.~Anderhub, L.~Antonelli, P.~Antoranz, M.~Backes et~al.,
  \emph{Very high energy gamma-ray observations of strong flaring activity in
  m87 in 2008 february}, {\emph{The Astrophysical Journal Letters} {\bf 685}
  (2008) L23}.

\bibitem{aliu2012}
E.~Aliu, T.~Arlen, T.~Aune, M.~Beilicke, W.~Benbow, A.~Bouvier et~al.,
  \emph{Veritas observations of day-scale flaring of m 87 in 2010 april},
  {\emph{The Astrophysical Journal} {\bf 746} (2012) 141}.

\bibitem{Murase:2011yw}
K.~Murase, \emph{{High-Energy Emission Induced by Ultra-high-Energy Photons as
  a Probe of Ultra-high-Energy Cosmic-Ray Accelerators Embedded in the Cosmic
  Web}}, \href{http://dx.doi.org/10.1088/2041-8205/745/2/L16}{\emph{Astrophys.
  J.} {\bf 745} (2012) L16}, [\href{http://arxiv.org/abs/1111.0936}{{\tt
  1111.0936}}].

\bibitem{Murase:2012xs}
K.~Murase and J.~F. Beacom, \emph{{Constraining Very Heavy Dark Matter Using
  Diffuse Backgrounds of Neutrinos and Cascaded Gamma Rays}},
  \href{http://dx.doi.org/10.1088/1475-7516/2012/10/043}{\emph{JCAP} {\bf 1210}
  (2012) 043}, [\href{http://arxiv.org/abs/1206.2595}{{\tt 1206.2595}}].

\bibitem{Murase:2011cy}
K.~Murase, C.~D. Dermer, H.~Takami and G.~Migliori, \emph{{Blazars as
  Ultra-High-Energy Cosmic-Ray Sources: Implications for TeV Gamma-Ray
  Observations}},
  \href{http://dx.doi.org/10.1088/0004-637X/749/1/63}{\emph{Astrophys. J.} {\bf
  749} (2012) 63}, [\href{http://arxiv.org/abs/1107.5576}{{\tt 1107.5576}}].

\bibitem{Murase:2012df}
K.~Murase, J.~F. Beacom and H.~Takami, \emph{{Gamma-Ray and Neutrino
  Backgrounds as Probes of the High-Energy Universe: Hints of Cascades, General
  Constraints, and Implications for TeV Searches}},
  \href{http://dx.doi.org/10.1088/1475-7516/2012/08/030}{\emph{JCAP} {\bf 1208}
  (2012) 030}, [\href{http://arxiv.org/abs/1205.5755}{{\tt 1205.5755}}].

\bibitem{Berezinsky:2016feh}
V.~Berezinsky and O.~Kalashev, \emph{{High energy electromagnetic cascades in
  extragalactic space: physics and features}},
  \href{http://dx.doi.org/10.1103/PhysRevD.94.023007}{\emph{Phys. Rev.} {\bf
  D94} (2016) 023007}, [\href{http://arxiv.org/abs/1603.03989}{{\tt
  1603.03989}}].

\bibitem{aharonian1983}
F.~Aharonian, A.~Atoian and A.~Nagapetian, \emph{Photoproduction of
  electron-positron pairs in compact x-ray sources}, {\emph{Astrofizika} {\bf
  19} (1983) 323--334}.

\bibitem{aharonian1981}
F.~Aharonian and A.~Atoyan, \emph{Compton scattering of relativistic electrons
  in compact x-ray sources}, {\emph{Astrophysics and Space Science} {\bf 79}
  (1981) 321--336}.

\bibitem{schlickeiser2010}
R.~Schlickeiser and J.~Ruppel, \emph{Klein--nishina steps in the energy
  spectrum of galactic cosmic-ray electrons}, {\emph{New Journal of Physics}
  {\bf 12} (2010) 033044}.

\bibitem{Moskalenko2006}
I.~V. Moskalenko, T.~A. Porter and A.~W. Strong, \emph{Attenuation of very high
  energy gamma rays by the milky way interstellar radiation field}, {\emph{The
  Astrophysical Journal Letters} {\bf 640} (2006) L155}.

\bibitem{Zhang2006}
J.-L. Zhang, X.-J. Bi and H.-B. Hu, \emph{Very high-energy gamma ray absorption
  by the galactic interstellar radiation field}, {\emph{Astronomy \&
  Astrophysics} {\bf 449} (2006) 641--643}.

\bibitem{Porter2005}
T.~Porter and A.~Strong, \emph{A new estimate of the galactic interstellar
  radiation field between 0.1 microns and 1000 microns}, {\emph{arXiv preprint
  astro-ph/0507119} (2005) }.

\bibitem{Dominguez:2010bv}
A.~Dominguez et~al., \emph{{Extragalactic Background Light Inferred from AEGIS
  Galaxy SED-type Fractions}},
  \href{http://dx.doi.org/10.1111/j.1365-2966.2010.17631.x}{\emph{Mon. Not.
  Roy. Astron. Soc.} {\bf 410} (2011) 2556},
  [\href{http://arxiv.org/abs/1007.1459}{{\tt 1007.1459}}].

\bibitem{Falcone:2010fk}
{\scshape Fermi} collaboration, A.~A. Abdo et~al., \emph{{Fermi Large Area
  Telescope View of the Core of the Radio Galaxy Centaurus A}},
  \href{http://dx.doi.org/10.1088/0004-637X/719/2/1433}{\emph{Astrophys. J.}
  {\bf 719} (2010) 1433--1444}, [\href{http://arxiv.org/abs/1006.5463}{{\tt
  1006.5463}}].

\bibitem{Aartsen:2016xlq}
{\scshape IceCube} collaboration, M.~G. Aartsen et~al., \emph{{Observation and
  Characterization of a Cosmic Muon Neutrino Flux from the Northern Hemisphere
  using six years of IceCube data}},
  \href{http://dx.doi.org/10.3847/0004-637X/833/1/3}{\emph{Astrophys. J.} {\bf
  833} (2016) 3}, [\href{http://arxiv.org/abs/1607.08006}{{\tt 1607.08006}}].

\bibitem{Aartsen:2015rwa}
{\scshape IceCube} collaboration, M.~G. Aartsen et~al., \emph{{Evidence for
  Astrophysical Muon Neutrinos from the Northern Sky with IceCube}},
  \href{http://dx.doi.org/10.1103/PhysRevLett.115.081102}{\emph{Phys. Rev.
  Lett.} {\bf 115} (2015) 081102}, [\href{http://arxiv.org/abs/1507.04005}{{\tt
  1507.04005}}].

\bibitem{Aartsen:2015knd}
{\scshape IceCube} collaboration, M.~G. Aartsen et~al., \emph{{A combined
  maximum-likelihood analysis of the high-energy astrophysical neutrino flux
  measured with IceCube}},
  \href{http://dx.doi.org/10.1088/0004-637X/809/1/98}{\emph{Astrophys. J.} {\bf
  809} (2015) 98}, [\href{http://arxiv.org/abs/1507.03991}{{\tt 1507.03991}}].

\bibitem{Neronov:2009gh}
A.~Neronov and D.~V. Semikoz, \emph{{Sensitivity of gamma-ray telescopes for
  detection of magnetic fields in intergalactic medium}},
  \href{http://dx.doi.org/10.1103/PhysRevD.80.123012}{\emph{Phys. Rev.} {\bf
  D80} (2009) 123012}, [\href{http://arxiv.org/abs/0910.1920}{{\tt
  0910.1920}}].

\bibitem{Tashiro:2013bxa}
H.~Tashiro and T.~Vachaspati, \emph{{Cosmological magnetic field correlators
  from blazar induced cascade}},
  \href{http://dx.doi.org/10.1103/PhysRevD.87.123527}{\emph{Phys. Rev.} {\bf
  D87} (2013) 123527}, [\href{http://arxiv.org/abs/1305.0181}{{\tt
  1305.0181}}].

\bibitem{Chen:2014rsa}
W.~Chen, J.~H. Buckley and F.~Ferrer, \emph{{Search for GeV γ -Ray Pair Halos
  Around Low Redshift Blazars}},
  \href{http://dx.doi.org/10.1103/PhysRevLett.115.211103}{\emph{Phys. Rev.
  Lett.} {\bf 115} (2015) 211103}, [\href{http://arxiv.org/abs/1410.7717}{{\tt
  1410.7717}}].

\bibitem{Caprioli:2015zka}
D.~Caprioli, \emph{{"Espresso" Acceleration of Ultra-high-energy Cosmic Rays}},
  \href{http://dx.doi.org/10.1088/2041-8205/811/2/L38}{\emph{Astrophys. J.}
  {\bf 811} (2015) L38}, [\href{http://arxiv.org/abs/1505.06739}{{\tt
  1505.06739}}].

\bibitem{Simet:2009ne}
M.~Simet and D.~Hooper, \emph{{Astrophysical Uncertainties in the Cosmic Ray
  Electron and Positron Spectrum From Annihilating Dark Matter}},
  \href{http://dx.doi.org/10.1088/1475-7516/2009/08/003}{\emph{JCAP} {\bf 0908}
  (2009) 003}, [\href{http://arxiv.org/abs/0904.2398}{{\tt 0904.2398}}].

\bibitem{Trotta:2010mx}
R.~Trotta, G.~Johannesson, I.~V. Moskalenko, T.~A. Porter, R.~R. de~Austri and
  A.~W. Strong, \emph{{Constraints on cosmic-ray propagation models from a
  global Bayesian analysis}},
  \href{http://dx.doi.org/10.1088/0004-637X/729/2/106}{\emph{Astrophys. J.}
  {\bf 729} (2011) 106}, [\href{http://arxiv.org/abs/1011.0037}{{\tt
  1011.0037}}].

\bibitem{Churazov:2003hr}
E.~Churazov, W.~Forman, C.~Jones and H.~Bohringer, \emph{{Xmm-newton
  observations of the perseus cluster I: the temperature and surface brightness
  structure}}, \href{http://dx.doi.org/10.1086/374923}{\emph{Astrophys. J.}
  {\bf 590} (2003) 225--237},
  [\href{http://arxiv.org/abs/astro-ph/0301482}{{\tt astro-ph/0301482}}].

\bibitem{1986ApJ...310..637T}
G.~{Trinchieri}, G.~{Fabbiano} and C.~R. {Canizares}, \emph{{The X-ray surface
  brightness distribution and spectral properties of six early-type galaxies}},
  \href{http://dx.doi.org/10.1086/164716}{\emph{\apj} {\bf 310} (Nov., 1986)
  637--659}.

\bibitem{Kraft:2003gp}
R.~P. Kraft, S.~Vazquez, W.~R. Forman, C.~Jones, S.~S. Murray, M.~J. Hardcastle
  et~al., \emph{{X-ray emission from the hot ISM and SW radio lobe of the
  nearby radio galaxy Centaurus A}},
  \href{http://dx.doi.org/10.1086/375533}{\emph{Astrophys. J.} {\bf 592} (2003)
  129--146}, [\href{http://arxiv.org/abs/astro-ph/0304363}{{\tt
  astro-ph/0304363}}].

\bibitem{Allen:2006mh}
S.~W. Allen, R.~J.~H. Dunn, A.~C. Fabian, G.~B. Taylor and C.~S. Reynolds,
  \emph{{The relation between accretion rate and jet power in x-ray luminous
  elliptical galaxies}},
  \href{http://dx.doi.org/10.1111/j.1365-2966.2006.10778.x}{\emph{Mon. Not.
  Roy. Astron. Soc.} {\bf 372} (2006) 21--30},
  [\href{http://arxiv.org/abs/astro-ph/0602549}{{\tt astro-ph/0602549}}].

\bibitem{Fukazawa:2005nx}
Y.~Fukazawa, J.~G. Betoya-Nonesa, J.~Pu, A.~Ohto and N.~Kawano, \emph{{Scaling
  mass profiles around elliptical galaxies observed with chandra and
  xmm-newton}}, \href{http://dx.doi.org/10.1086/498081}{\emph{Astrophys. J.}
  {\bf 636} (2006) 698--711},
  [\href{http://arxiv.org/abs/astro-ph/0509521}{{\tt astro-ph/0509521}}].

\bibitem{Taylor:2001jw}
G.~B. Taylor, A.~C. Fabian and S.~W. Allen, \emph{{Magnetic fields in the
  centaurus cluster}},
  \href{http://dx.doi.org/10.1046/j.1365-8711.2002.05555.x}{\emph{Mon. Not.
  Roy. Astron. Soc.} {\bf 334} (2002) 769},
  [\href{http://arxiv.org/abs/astro-ph/0109337}{{\tt astro-ph/0109337}}].

\bibitem{Kim:1997dv}
D.-W. Kim, G.~Fabbiano and G.~Mackie, \emph{{Rosat x-ray observations of the
  radio galaxy ngc 1316 (fornax a)}},
  \href{http://dx.doi.org/10.1086/305476}{\emph{Astrophys. J.} {\bf 497} (1998)
  699}, [\href{http://arxiv.org/abs/astro-ph/9711098}{{\tt astro-ph/9711098}}].

\bibitem{Owen:2000vi}
F.~N. Owen, J.~a. Eilek and N.~E. Kassim, \emph{{M87 at 90cm: a different
  picture}}, \href{http://dx.doi.org/10.1086/317151}{\emph{Astrophys. J.} {\bf
  543} (2000) 611}, [\href{http://arxiv.org/abs/astro-ph/0006150}{{\tt
  astro-ph/0006150}}].

\bibitem{Aloisio:2004jda}
R.~Aloisio and V.~Berezinsky, \emph{{Diffusive propagation of UHECR and the
  propagation theorem}},
  \href{http://dx.doi.org/10.1086/421869}{\emph{Astrophys. J.} {\bf 612} (2004)
  900--913}, [\href{http://arxiv.org/abs/astro-ph/0403095}{{\tt
  astro-ph/0403095}}].

\bibitem{1994A&A...286..983M}
K.~{Mannheim} and R.~{Schlickeiser}, \emph{{Interactions of cosmic ray
  nuclei}}, {\emph{\aap} {\bf 286} (June, 1994) },
  [\href{http://arxiv.org/abs/astro-ph/9402042}{{\tt astro-ph/9402042}}].

\bibitem{Taylor:2013gga}
A.~M. Taylor, \emph{{UHECR Composition Models}},
  \href{http://dx.doi.org/10.1016/j.astropartphys.2013.11.006}{\emph{Astropart.
  Phys.} {\bf 54} (2014) 48--53}, [\href{http://arxiv.org/abs/1401.0199}{{\tt
  1401.0199}}].

\bibitem{Taylor:2011ta}
A.~M. Taylor, M.~Ahlers and F.~A. Aharonian, \emph{{The need for a local source
  of UHE CR nuclei}},
  \href{http://dx.doi.org/10.1103/PhysRevD.84.105007}{\emph{Phys. Rev.} {\bf
  D84} (2011) 105007}, [\href{http://arxiv.org/abs/1107.2055}{{\tt
  1107.2055}}].

\bibitem{Beckmann:2011rh}
V.~Beckmann, P.~Jean, P.~Lubinski, S.~Soldi and R.~Terrier, \emph{{The Hard
  X-ray Emission of Cen A}},
  \href{http://dx.doi.org/10.1051/0004-6361/201016020}{\emph{Astron.
  Astrophys.} {\bf 531} (2011) A70},
  [\href{http://arxiv.org/abs/1104.4253}{{\tt 1104.4253}}].

\bibitem{Abreu:2010ab}
{\scshape Pierre Auger} collaboration, P.~Abreu et~al., \emph{{Update on the
  correlation of the highest energy cosmic rays with nearby extragalactic
  matter}},
  \href{http://dx.doi.org/10.1016/j.astropartphys.2010.08.010}{\emph{Astropart.
  Phys.} {\bf 34} (2010) 314--326}, [\href{http://arxiv.org/abs/1009.1855}{{\tt
  1009.1855}}].

\bibitem{Liu:2012sq}
R.-Y. Liu, X.-Y. Wang, W.~Wang and A.~M. Taylor, \emph{{On the excess of
  ultra-high energy cosmic rays in the direction of Centaurus A}},
  \href{http://dx.doi.org/10.1088/0004-637X/755/2/139}{\emph{Astrophys. J.}
  {\bf 755} (2012) 139}, [\href{http://arxiv.org/abs/1206.3907}{{\tt
  1206.3907}}].

\bibitem{Biermann:2011wf}
P.~L. Biermann and V.~de~Souza, \emph{{Centaurus A: the one extragalactic
  source of cosmic rays with energies above the knee}},
  \href{http://dx.doi.org/10.1088/0004-637X/746/1/72}{\emph{Astrophys. J.} {\bf
  746} (2012) 72}, [\href{http://arxiv.org/abs/1106.0625}{{\tt 1106.0625}}].

\bibitem{Yuksel:2012ee}
H.~Yuksel, T.~Stanev, M.~D. Kistler and P.~P. Kronberg, \emph{{The Centaurus A
  Ultrahigh-Energy Cosmic Ray Excess and the Local Extragalactic Magnetic
  Field}}, \href{http://dx.doi.org/10.1088/0004-637X/758/1/16}{\emph{Astrophys.
  J.} {\bf 758} (2012) 16}, [\href{http://arxiv.org/abs/1203.3197}{{\tt
  1203.3197}}].

\bibitem{Kim:2012rp}
H.~B. Kim, \emph{{Centaurus A as a point source of Ultra-High Energy Cosmic
  Rays}}, \href{http://dx.doi.org/10.1088/0004-637X/764/2/121}{\emph{Astrophys.
  J.} {\bf 764} (2013) 121}, [\href{http://arxiv.org/abs/1206.3839}{{\tt
  1206.3839}}].

\bibitem{Farrar:2012gm}
G.~R. Farrar, R.~Jansson, I.~J. Feain and B.~M. Gaensler, \emph{{Galactic
  magnetic deflections and Centaurus A as a UHECR source}},
  \href{http://dx.doi.org/10.1088/1475-7516/2013/01/023}{\emph{JCAP} {\bf 1301}
  (2013) 023}, [\href{http://arxiv.org/abs/1211.7086}{{\tt 1211.7086}}].

\bibitem{Sushchov:2012ri}
O.~Sushchov, O.~Kobzar, B.~Hnatyk and V.~Marchenko, \emph{{Search for ultra
  high energy cosmic ray sources. FRI radio galaxy Centaurus A}},
  \href{http://dx.doi.org/10.3103/S0884591312060062}{\emph{Kinematics Phys.
  Celest. Bodies} {\bf 28} (2012) 270},
  [\href{http://arxiv.org/abs/1212.1402}{{\tt 1212.1402}}].

\bibitem{Keivani:2014kua}
A.~Keivani, G.~R. Farrar and M.~Sutherland, \emph{{Magnetic Deflections of
  Ultra-High Energy Cosmic Rays from Centaurus A}},
  \href{http://dx.doi.org/10.1016/j.astropartphys.2014.07.001}{\emph{Astropart.
  Phys.} {\bf 61} (2014) 47--55}, [\href{http://arxiv.org/abs/1406.5249}{{\tt
  1406.5249}}].

\bibitem{Aartsen:2016oji}
{\scshape IceCube} collaboration, M.~G. Aartsen et~al., \emph{{All-sky Search
  for Time-integrated Neutrino Emission from Astrophysical Sources with 7 yr of
  IceCube Data}},
  \href{http://dx.doi.org/10.3847/1538-4357/835/2/151}{\emph{Astrophys. J.}
  {\bf 835} (2017) 151}, [\href{http://arxiv.org/abs/1609.04981}{{\tt
  1609.04981}}].

\bibitem{Albert:2017ohr}
{\scshape ANTARES} collaboration, A.~Albert et~al., \emph{{First all-flavour
  Neutrino Point-like Source Search with the ANTARES Neutrino Telescope}},
  \href{http://arxiv.org/abs/1706.01857}{{\tt 1706.01857}}.

\bibitem{Aartsen:2017eiu}
{\scshape IceCube} collaboration, M.~G. Aartsen et~al., \emph{{Search for
  astrophysical sources of neutrinos using cascade events in IceCube}},
  \href{http://arxiv.org/abs/1705.02383}{{\tt 1705.02383}}.

\bibitem{aartsen2014searches}
M.~Aartsen, M.~Ackermann, J.~Adams, J.~Aguilar, M.~Ahlers, M.~Ahrens et~al.,
  \emph{Searches for extended and point-like neutrino sources with four years
  of icecube data}, {\emph{The Astrophysical Journal} {\bf 796} (2014) 109}.

\bibitem{adrian2016first}
S.~Adri{\'a}n-Mart{\'\i}nez, A.~Albert, M.~Andr{\'e}, G.~Anton, M.~Ardid, J.-J.
  Aubert et~al., \emph{The first combined search for neutrino point-sources in
  the southern hemisphere with the antares and icecube neutrino telescopes},
  {\emph{The Astrophysical Journal} {\bf 823} (2016) 65}.

\bibitem{Blaufuss:2015muc}
{\scshape IceCube-Gen2} collaboration, E.~Blaufuss, C.~Kopper and C.~Haack,
  \emph{{The IceCube-Gen2 High Energy Array}}, {\emph{PoS} {\bf ICRC2015}
  (2016) 1146}.

\bibitem{Adrian-Martinez:2016fdl}
{\scshape KM3Net} collaboration, S.~Adrian-Martinez et~al., \emph{{Letter of
  intent for KM3NeT 2.0}},
  \href{http://dx.doi.org/10.1088/0954-3899/43/8/084001}{\emph{J. Phys.} {\bf
  G43} (2016) 084001}, [\href{http://arxiv.org/abs/1601.07459}{{\tt
  1601.07459}}].

\bibitem{Halzen:2003fi}
F.~Halzen and D.~Hooper, \emph{{IceCube-Plus: An Ultrahigh-energy neutrino
  telescope}},
  \href{http://dx.doi.org/10.1088/1475-7516/2004/01/002}{\emph{JCAP} {\bf 0401}
  (2004) 002}, [\href{http://arxiv.org/abs/astro-ph/0310152}{{\tt
  astro-ph/0310152}}].

\end{thebibliography}\endgroup
 \bibliographystyle{JHEP}

\end{document}